\documentclass[12pt,a4paper]{article}
\pdfoutput=1
\usepackage{amsmath}   
\usepackage{slashed}   
\usepackage{subfigure}   

\usepackage[american]{babel}
\usepackage[dvips]{graphicx}
\usepackage{bbm}
\usepackage{color}
\usepackage[unicode]{hyperref}

\newcommand{\be}{\begin{equation}}   
\newcommand{\ee}{\end{equation}}   
\newcommand{\beq}{\begin{equation}}   
\newcommand{\eeq}{\end{equation}}   
\newcommand{\bea}{\begin{eqnarray}}   
\newcommand{\eea}{\end{eqnarray}}   
\newcommand{\nn}{\nonumber}   
   
\numberwithin{equation}{section}
\begin{document}   
\baselineskip=15.5pt   
\pagestyle{plain}   
\setcounter{page}{1}   
\def\del{{\partial}}   
\def\vev#1{\left\langle #1 \right\rangle}   
\def\cn{{\cal N}}   
\def\co{{\cal O}}   
\newfont{\Bbb}{msbm10 scaled 1200}     
\newcommand{\mathbb}[1]{\mbox{\Bbb #1}}   
\def\IC{{\mathbb C}}   
\def\IR{{\mathbb R}}   
\def\IZ{{\mathbb Z}}   
\def\RP{{\bf RP}}   
\def\CP{{\bf CP}}   
\def\Poincare{{Poincar\'e }}   
\def\tr{{\rm tr}}   
\def\tp{{\tilde \Phi}}   
\def\TL{\hfil$\displaystyle{##}$}   
\def\TR{$\displaystyle{{}##}$\hfil}   
\def\TC{\hfil$\displaystyle{##}$\hfil}   
\def\TT{\hbox{##}}   
\def\HLINE{\noalign{\vskip1\jot}\hline\noalign{\vskip1\jot}}   
\def\seqalign#1#2{\vcenter{\openup1\jot   
  \halign{\strut #1\cr #2 \cr}}}   
\def\lbldef#1#2{\expandafter\gdef\csname #1\endcsname {#2}}   
\def\eqn#1#2{\lbldef{#1}{(\ref{#1})}%
\begin{equation} #2 \label{#1} \end{equation}}   
\def\eqalign#1{\vcenter{\openup1\jot   
    \halign{\strut\span\TL & \span\TR\cr #1 \cr   
   }}}   
\def\eno#1{(\ref{#1})}   
\def\href#1#2{#2}   
\def\half{{1 \over 2}}   
\def\ads{{\it AdS}}   
\def\adsp{{\it AdS}$_{p+2}$}   
\def\cft{{\it CFT}}   
\newcommand{\ber}{\begin{eqnarray}}   
\newcommand{\eer}{\end{eqnarray}}   
\newcommand{\beqar}{\begin{eqnarray}}   
\newcommand{\cN}{{\cal N}}   
\newcommand{\cO}{{\cal O}}   
\newcommand{\cA}{{\cal A}}   
\newcommand{\cT}{{\cal T}}   
\newcommand{\cF}{{\cal F}}   
\newcommand{\cC}{{\cal C}}   
\newcommand{\cR}{{\cal R}}   
\newcommand{\cW}{{\cal W}}   
\newcommand{\eeqar}{\end{eqnarray}}   
\newcommand{\tht}{\thteta}   
\newcommand{\lm}{\lambda}\newcommand{\Lm}{\Lambda}   
\newcommand{\eps}{\epsilon}   
\newcommand{\nonu}{\nonumber}   
\newcommand{\oh}{\displaystyle{\frac{1}{2}}}   
\newcommand{\dsl}   
  {\kern.06em\hbox{\raise.15ex\hbox{$/$}\kern-.56em\hbox{$\partial$}}}   
\newcommand{\id}{i\!\!\not\!\partial}   
\newcommand{\as}{\not\!\! A}   
\newcommand{\ps}{\not\! p}   
\newcommand{\ks}{\not\! k}   
\newcommand{\D}{{\cal{D}}}   
\newcommand{\dv}{d^2x}   
\newcommand{\Z}{{\cal Z}}   
\newcommand{\N}{{\cal N}}   
\newcommand{\Dsl}{\not\!\! D}   
\newcommand{\Bsl}{\not\!\! B}   
\newcommand{\Psl}{\not\!\! P}   
\newcommand{\eeqarr}{\end{eqnarray}}   
\newcommand{\ZZ}{{\rm \kern 0.275em Z \kern -0.92em Z}\;}   
                                                                                                       
\def\del{{\delta^{\hbox{\sevenrm B}}}} \def\ex{{\hbox{\rm e}}}   
\def\azb{A_{\bar z}} \def\az{A_z} \def\bzb{B_{\bar z}} \def\bz{B_z}   
\def\czb{C_{\bar z}} \def\cz{C_z} \def\dzb{D_{\bar z}} \def\dz{D_z}   
\def\im{{\hbox{\rm Im}}} \def\mod{{\hbox{\rm mod}}} \def\tr{{\hbox{\rm Tr}}}   
\def\ch{{\hbox{\rm ch}}} \def\imp{{\hbox{\sevenrm Im}}}   
\def\trp{{\hbox{\sevenrm Tr}}} \def\vol{{\hbox{\rm Vol}}}   
\def\rl{\Lambda_{\hbox{\sevenrm R}}} \def\wl{\Lambda_{\hbox{\sevenrm W}}}   
\def\fc{{\cal F}_{k+\cox}} \def\vev{vacuum expectation value}   
\def\nodiv{\mid{\hbox{\hskip-7.8pt/}}}   
\def\ie{{\em i.e.}}   
\def\ie{\hbox{\it i.e.}}

\def\CC{{\mathchoice   
{\rm C\mkern-8mu\vrule height1.45ex depth-.05ex   
width.05em\mkern9mu\kern-.05em}   
{\rm C\mkern-8mu\vrule height1.45ex depth-.05ex   
width.05em\mkern9mu\kern-.05em}   
{\rm C\mkern-8mu\vrule height1ex depth-.07ex   
width.035em\mkern9mu\kern-.035em}   
{\rm C\mkern-8mu\vrule height.65ex depth-.1ex   
width.025em\mkern8mu\kern-.025em}}}   
                                                                                                       
\def\RR{{\rm I\kern-1.6pt {\rm R}}}   
\def\NN{{\rm I\!N}}   
\def\ZZ{{\rm Z}\kern-3.8pt {\rm Z} \kern2pt}   
\def\IB{\relax{\rm I\kern-.18em B}}   
\def\ID{\relax{\rm I\kern-.18em D}}   
\def\II{\relax{\rm I\kern-.18em I}}   
\def\IP{\relax{\rm I\kern-.18em P}}   
\newcommand{\CS}{{\scriptstyle {\rm CS}}}   
\newcommand{\CSs}{{\scriptscriptstyle {\rm CS}}}   
\newcommand{\rc}{\nonumber\\}   
\newcommand{\bear}{\begin{eqnarray}}   
\newcommand{\eear}{\end{eqnarray}}   
\newcommand{\W}{{\cal W}}   
\newcommand{\F}{{\cal F}}   
\newcommand{\x}{{\cal O}}   
\newcommand{\LL}{{\cal L}}   
                                                                                                       
\def\mani{{\cal M}}   
\def\calo{{\cal O}}   
\def\calb{{\cal B}}   
\def\calw{{\cal W}}   
\def\calz{{\cal Z}}   
\def\cald{{\cal D}}   
\def\calc{{\cal C}}   
\def\to{\rightarrow}   
\def\ele{{\hbox{\sevenrm L}}}   
\def\ere{{\hbox{\sevenrm R}}}   
\def\zb{{\bar z}}   
\def\wb{{\bar w}}   
\def\nodiv{\mid{\hbox{\hskip-7.8pt/}}}   
\def\menos{\hbox{\hskip-2.9pt}}   
\def\dr{\dot R_}   
\def\drr{\dot r_}   
\def\ds{\dot s_}   
\def\da{\dot A_}   
\def\dga{\dot \gamma_}   
\def\ga{\gamma_}   
\def\dal{\dot\alpha_}   
\def\al{\alpha_}   
\def\cl{{closed}}   
\def\cls{{closing}}   
\def\vev{vacuum expectation value}   
\def\tr{{\rm Tr}}   
\def\to{\rightarrow}   
\def\too{\longrightarrow}   
\def\a{\alpha}   
\def\b{\beta}   
\def\c{\gamma}   
\def\d{\delta}   
\def\e{\epsilon}           
\def\f{\phi}               
\def\vf{\varphi}  \def\tvf{\tilde{\varphi}}   
\def\vp{\varphi}   
\def\g{\gamma}   
\def\h{\eta}   
\def\i{\iota}   
\def\j{\psi}   
\def\k{\kappa}                    
\def\l{\lambda}   
\def\m{\mu}   
\def\n{\nu}   
\def\o{\omega}  \def\w{\omega}   
\def\q{\theta}  \def\th{\theta}                  
\def\r{\rho}                                     
\def\s{\sigma}                                   
\def\t{\tau}   
\def\u{\upsilon}   
\def\x{\xi}   
\def\z{\zeta}   
\def\pt{\tilde{\varphi}}   
\def\tt{\tilde{\theta}}   
\def\lab{\label}     
\def\6{\partial}   
\def\wg{\wedge}   
\def\atanh{{\rm arctanh}}   
\def\bpsi{\bar{\psi}}   
\def\bt{\bar{\theta}}   
\def\bvf{\bar{\varphi}}   
%
\newfont{\namefont}{cmr10}   
\newfont{\addfont}{cmti7 scaled 1440}   
\newfont{\boldmathfont}{cmbx10}   
\newfont{\headfontb}{cmbx10 scaled 1728}   

\def\red{\textcolor[rgb]{0.98,0.00,0.00}}
\def\green{\textcolor[rgb]{0,.3, 1}}

\begin{titlepage}

\vspace{0.1in}
\begin{flushright}
  UTTG-38-13\\
 FPAUO 14/02
 \end{flushright}

\begin{center}
\Large \bf  New Type IIB Backgrounds and Aspects of Their Field Theory Duals.
\end{center}
\vskip 0.2truein
\begin{center}
Elena Caceres$^{a,b,}$\footnote{elenac@zippy.ph.utexas.edu}, 
Niall T. Macpherson $^{c,d,}$\footnote{pymacpherson@swansea.ac.uk} and 
Carlos N\'u\~nez$^{c,e,}$\footnote{c.nunez@swansea.ac.uk} 
\\\vspace{.2in}
{\it $a$: Facultad de Ciencias, Universidad de Colima\\ 
Bernal Diaz del Castillo 340, Colima, Mexico }\\
\vspace{.2in}
{\it $b$: Theory Group, Department of Physics, 
The University of Texas at Austin\\
	Austin, TX 78712, USA}\\
\vspace{0.2in}
{\it $c$: Department of Physics, Swansea University\\
 Singleton Park, Swansea SA2 8PP, United Kingdom.}\\
\vspace{0.2in}
{\it $d$:Department of Physics, University of Oviedo.\\Avda Calvo Sotelo 18, 33007 Oviedo, Spain.}\\
\vspace{0.2in}
{\it $e$: CP3-Origins and DIAS. University of Southern Denmark.}\\
\vspace{0.2in}

\vskip 5mm

\vspace{0.2in}
\end{center}
\vspace{0.2in}
\centerline{\bf Abstract:}

In this paper we study aspects of geometries in Type IIA and Type IIB
String theory and elaborate on their field theory dual pairs. The 
backgrounds are associated with reductions to Type IIA of solutions 
with $G_2$ holonomy in eleven dimensions. 
We classify these backgrounds according to their
G-structure, perform a non-Abelian T-duality on them and find new Type IIB
configurations presenting {\it dynamical} $SU(2)$-structure. 
We study some aspects of the associated field theories 
defined by these new backgrounds. 
Various technical details are clearly spelled out.
\end{titlepage}
\setcounter{footnote}{0}

\tableofcontents

\setcounter{footnote}{0}
\renewcommand{\theequation}{{\rm\thesection.\arabic{equation}}}
\section{Introduction and General Idea of this Paper.}   
The Maldacena conjecture \cite{Maldacena:1997re},    
\cite{Witten:1998qj} substantially changed the panorama of theoretical physics.   
In the last fourteen years, the area has been dominated by  ideas tightly associated with    
gauge-Strings dualities.  In most of the examples,    
the idea is to use dualities   
to study interesting   aspects of quantum field theories (QFTs) which cannot be approached by  perturbative techniques.    
A  massive amount of work  deals with     
theories with minimal  SUSY in different number of dimensions.    
This lead to the discovery    
of new string backgrounds \cite{Klebanov:2000hb} 
that encode phenomena as diverse and  nontrivial   
as confinement, breaking of global symmetries, presence of domain-wall like objects,    
diverse correlation functions, etc. In all these cases the QFT is strongly coupled but the duality relates these nontrivial phenomena to semi classical 
calculations on the string theory side.   This line of work evolved in many directions,    
with many applications to different branches of Theoretical Physics.    
   
One of these directions is  the  construction of duals to field    
theories in four dimensions realized on the worldvolume of $D_{p>3}$ branes,    
where ($p-3$) directions have been compactified on a small manifold. The compactification is (usually) performed  
in a way that preserves the smallest amount of SUSY.    
These field theories, are higher-dimensional in 
disguise; the whole construction is 
in spirit,  similar  to the Kaluza-Klein idea. 
In this paper, we will deal with    
one such example. We will consider  the case in which $D6$ branes   
wrap a calibrated three-cycle inside the deformed conifold.    
Extensions of this case to different number of dimensions,    
different number of preserved SUSY, etc; have been studied.     
In particular, if these configurations in Type IIA string theory are    
lifted to eleven dimensions, the configurations become purely geometric,    
leading to the associated seven-dimensional   
spaces possesing $G_2$ holonomy.    
This line of research \cite{Atiyah:2000zz}-\cite{Brandhuber:2001yi},    
was quite fertile, specially on the mathematical side where it lead to the construction of  new metrics with $G_2$ holonomy. However, it  did not give as many physically interesting result   
as its Type IIB counterparts \cite{Klebanov:2000hb}.    
In this work we  present a family of those `old'    
$G_2$ metrics, reduce the system to Type IIA and study some of its physical implications,    
making sharper the reasons for which they failed to capture some    
of the phenomena their Type IIB counterpart were able to calculate.   
   
In parallel with these `physically motivated' discoveries, a powerful    
line of research was developed, aiming to a complete classification of    
different backgrounds by specifying  their G-structure 
\cite{Gauntlett:2002sc}-\cite{Grana:2004bg}.    
In particular, in these four dimensional and SUSY preserving examples,    
it is possible to encode {\it all} the information about the background    
(BPS equations, metric, fluxes, calibrated sub-manifolds, etc),    
in a set of forms defined on the space `external' to the Minkowski coordinates.   
Furthermore, the  $SU(2)$ and $SU(3)$ structures typical of these backgrounds, their associated    
pure spinors and forms  encode in subtle ways quite common operations    
in QFT  \cite{Maldacena:2009mw}.   
In this paper we  complement the above mentioned    
study of the type IIA backgrounds associated with the wrapped D6 branes   
and their precise description in terms of G-structures.   
   
We  also perform an operation on the geometry    
called non-Abelian T-duality. 
For a sample of old and recent research on the topic, see    
\cite{de la Ossa:1992vc}-
\cite{Sfetsos:2013wia}.
We    
generate new Type IIB solutions    that preserve
four supercharges; hence it is dual to a minimally SUSY
4-d QFT . We describe the result of
the non-Abelian    
T-duality in terms of the generated G-structure.    
We believe, ours is  one of the first few examples 
of {\it dynamical} $SU(2)$-structure in Type IIB.
We will use the word 'dynamical' to denote the fact that the quantities
$k_{\perp}, k_{\|}$ defined in eq.(\ref{useful}), 
are point dependent, changing value through out the internal manifold.    
We will propose a relation between 
the `dynamical' character of the $SU(2)$-structure    
and the phenomena of confinement in the dual QFT.   
   
The structure of the paper is the following. In    
Section \ref{presentationofbackground}---that contains a fair amount of review   
but also various original pieces,    
we will sumarise the eleven-dimensional    
and Type IIA supergravity solutions    
that will act as the `seed backgrounds' for our    
non-Abelian T-duality generating technique.   
Their G-structure will be carefully discussed.    
We will also present the explicit numerical solutions    
to the BPS equations and clarify their asymptotics.   
In Section \ref{NATDx},   
the action of non-Abelian T-duality on the Type IIA backgrounds,    
 the new generated  solutions in Type IIB  and   a  discussion of their 
G-structure will be spelled-out in detail.   
Different dual field theory aspects of the original   
and of the generated solution will be described 
in Section \ref{QFTaspects}.    
Finally, we close the paper in Section \ref{conclusionsxx}   
with some global remarks and proposing    
topics to be investigated. An appendix that discusses
the delicate numerical study, complements the presentation.

\section{Presentation of the Background.} \label{presentationofbackground}   
We will start with the pure metric configuration in    
eleven-dimensions found in \cite{Brandhuber:2001kq}, \cite{Cvetic:2001kp}.   
We consider the family  called ${\cal D}_7$.    
The notation we will adopt is that of \cite{Cvetic:2001kp}.   
We will have two sets of left invariant forms of $SU(2)$,   
\bea   
\begin{array}{lcl}   
\sigma_1= \cos \psi_1\, d\theta + \sin \psi_1\, \sin \theta\, d\varphi &   
\quad , \quad &   
\Sigma_1= \cos {\psi_2}\, d{\tilde\theta} + \sin {\psi_2}\, \sin   
{\tilde\theta}\, d{\tilde\varphi} \\   
\sigma_2= -\sin \psi_1\, d\theta + \cos \psi_1\, \sin \theta\, d\varphi &   
\quad , \quad &   
\Sigma_2= -\sin {\psi_2}\, d{\tilde\theta} + \cos {\psi_2}\, \sin   
{\tilde\theta}\, d{\tilde\varphi} \\   
\sigma_3= d\psi_1 + \cos \theta\, d \varphi &   
\quad , \quad &   
\Sigma_3= d{\psi_2} + \cos {\tilde\theta}\, d{\tilde{\varphi}}   
\label{sigmas}   
\end{array}   
\eea   
which satisfy the $SU(2)$ algebras   
\be\label{sigmaalg}   
d\sigma_1 = - \sigma_2 \wedge \sigma_3~+ \mathrm{cyclic~ perms.},  \quad   
d\Sigma_1 = - \Sigma_2 \wedge \Sigma_3~+ \mathrm{cyclic~ perms.}   
\ee   
The eleven dimensional metric is of the form   
$   
ds_{11}^2= dx_{1,3}^2 +  ds_7^2,   
$   
with   
\be   
ds_7^2 = dr^2 + a^2\, [(\Sigma_1+ g\, \sigma_1)^2 +   
(\Sigma_2+ g\,\sigma_2)^2] + b^2\, (\sigma_1^2 + \sigma_2^2) +   
c^2 (\Sigma_3 +g_3\, \sigma_3)^2 + f^2 \sigma_3^2\,,\label{d7ans0}   
\ee   
where $a$, $b$, ${ c}$, ${ f}$, $g$ and ${ g}_3$ are functions only of the   
radial variable $r$.    
The six functions are not all independent, the relations   
\be   
g(r) = \frac{-a(r)f(r)}{2b(r)c(r)} , \;\; g_3(r) = -1 + 2g(r)^2 .   
\ee   
are necessary for the BPS system   
\bea\label{eq:Deqns}   
\dot{a} = -\frac{c}{2a} + \frac{a^5 f^2}{8 b^4 c^3}, & \;\; &   
\dot{b} = -\frac{c}{2b} - \frac{a^2 (a^2-3c^2)f^2}{8b^3c^3}, \nn \\   
\dot{c} = -1+\frac{c^2}{2 a^2}+\frac{c^2}{2 b^2}-\frac{3 a^2   
f^2}{8b^4}, & \;\; &   
\dot{f} = -\frac{a^4 f^3}{4 b^4 c^3},   
\eea   
to satisfy the equations of motion. We have checked that these equations   
imply that the eleven dimensional metric satisfies $R_{\mu\nu}=0$.    
\subsection{The Type IIA version.}\label{section: IIA}
For our purposes, we need the Type IIA    
version of the configuration presented above and we need to 
pick a $U(1)$ isometry to reduce on.   
The relevant $U(1)$ isometry is generated   
by the Killing vector $\partial_{\psi_1} + \partial_{\psi_2}$.   
Having this in mind we rewrite the metric in a way which   
makes the isometry manifest,   
\bea\label{flux2a}   
ds_{11}^2 & = &dx_{1,3}^2+ dr^2+ b^2 \left[ (\sigma_1)^2 + (\sigma_2)^2 \right] +   
a^2 \left[ (\Sigma_1 +g \sigma_1)^2 + (\Sigma_2 +g \sigma_2)^2\right]   
\nonumber \\   
& & + \frac{f^2 c^2}{f^2 + (1+g_3)^2 c^2}   
\left( \sigma_3 - \Sigma_3 \right)^2   
\nonumber \\   
 & & + \frac{1}{4} \left[ f^2 + (1+g_3)^2 c^2 \right]   
\left[ \sigma_3+\Sigma_3 + \frac{f^2-c^2(1-g_3^2)}{f^2+(1+g_3)^2 c^2}   
\left(\sigma_3-\Sigma_3 \right) \right]^2 ~,   
\eea   
Note that in this metric nothing depends on the combination    
$(\psi_2 + \psi_1)$.   
Now Kaluza-Klein reduction simply amounts to dropping the last   
line in eq.(\ref{flux2a}) which has been written as a complete square   
for that purpose.   
In particular we can now read off the dilaton    
and the RR one-form gauge field,   
\be\label{fluxdil}   
e^\phi = 2^{-3/2} \left[ f^2 + (1+g_3)^2 c^2 \right]^{3/4}~,\;\;\;\;   
A_1 = \frac{f^2-c^2(1-g_3^2)}{f^2+(1+g_3)^2 c^2}   
\left(\sigma_3-\Sigma_3 \right) + \cos \theta d\phi +   
\cos \tilde{\theta} d\tilde{\phi} ~.   
\ee   
The ten-dimensional metric in string frame is given by   
\bea\label{IIaflux}   
ds^2_{IIA} & = & \frac{1}{2}   
\Big\{ dx_{1,3}^2 +   
b^2 \left[ (\sigma_1)^2 + (\sigma_2)^2 \right] +   
a^2 \left[ (\Sigma_1 +g \sigma_1)^2 + (\Sigma_2 +g \sigma_2)^2\right]   
\nonumber \\   
& & + \frac{f^2 c^2}{f^2 + (1+g_3)^2 c^2}   
\left( \sigma_3 - \Sigma_3 \right)^2 + dr^2 \Big\} \times   
\left[ f^2 + (1+g_3)^2 c^2 \right]^{1/2}   
\eea   
Notice that the metric  depends explicitly  on    
$\psi=\psi_2-\psi_1$ and not on  the coordinate on which we reduced,    
 $\psi_+=\psi_2+\psi_1$. It is then 
advantageous to introduce a third set of one-forms:   
\beq   
\tilde\omega_1= \cos \psi\, d \tilde{\theta} + \sin \psi\, \sin \tilde{\theta}\, d\tilde{\varphi},~~~\tilde\omega_2= -\sin \psi\, d \tilde{\theta}    
+ \cos \psi\, \sin \tilde{\theta}\, d\tilde{\varphi},~~~\tilde\omega_3= d\psi + \cos \tilde{\theta}d\tilde{\varphi}.\;   
\eeq   
It should be pointed out here that the metric
is written in terms of two-pairs of 
left-invariant forms of $SU(2)$.
In the following section, we will perform a non-Abelian T-duality
on the $SU(2)$ described by the coordinates $(\theta,\varphi,\psi)$.

Upon rescaling the Minkowski part of the space 
by a constant $\mu$ and reinstating the factors of $\alpha', g_s$,    
the full metric, dilaton and RR field strength are
\footnote{One can send $ds^2_6\to A_1ds^2_6$, $F_2\to A_2 F_2$, $e^{-4\phi/3}\to A_3e^{-4\phi/3}$ 
and still have a solution of IIA supergravity, 
preserving $\mathcal{N}=1$ SUSY provided $A_1^2A_3^3=A_2^4$. 
We choose $A_1= \a'g_sN$, $A_2= \sqrt{\a}'g_sN$ and $A_3=(g_s N)^{2/3}$, 
so that the dilaton is independent of $\alpha'$. 
The parameter $\mu$ is just a scaling the $R^{1,3}$ coordinates.},     
\beq   
\begin{array}{ll}   
\vspace{3 mm}   
ds_{IIA,st}^2&=\alpha' g_s N e^{2A}\Bigg[\frac{\mu}{\alpha' g_sN} dx_{1,3}^2 + dr^2 + b^2(d\theta^2 + \sin^2\theta d\varphi^2)+\\   
\vspace{3 mm}   
&~~~~~~~a^2(\tilde\omega^1+gd\theta)^2+a^2(\tilde\omega^2+g\sin\theta d\varphi)^2+h^2(\tilde\omega^3-\cos\theta d\varphi)^2\Bigg]\\   
\vspace{3 mm}   
h^2&=\frac{c^2f^2}{f^2+c^2(1+g_3)^2},\;\;\;e^{4/3\phi}=\frac{c^2 f^2}{4 (g_s N)^{2/3} h^2},\;\;\;e^{4A}=\frac{c^2 f^2}{4 h^2}\\   
\frac{F_2}{\sqrt{\alpha'} g_s N}&=-(1+K)\sin\theta d\theta\wedge    
d\varphi +(K-1)\tilde\omega^1\wedge\tilde\omega^2-K'dr\wedge(\tilde\omega^3-\cos\theta d\varphi).\\   
\label{metricxxx}\end{array}   
\eeq   
where $$ K(r)= \frac{f^2 - c^2(1-g_3^2)}{f^2+ c^2(1+g_3)^2}.$$   
Note that $F_2$ contains two components with no    
`legs' on the radial coordinate $r$:   
\beq   
F_2\Bigg|_{r=r_0}= -Ng_s \sqrt{\alpha'}\Big[(K+1)\sin\theta d\theta\wedge 
d\varphi-(K-1)\sin\tilde{\theta} d\tilde{\theta}   
\wedge d\tilde{\varphi}\Big].   
\eeq   
Thus, we only have flux quantisation on cycles for    
which the $K(r)$ parts mutually cancel.    
For example on $\Sigma_2=[\tilde{\theta}=\theta$, $\tilde{\varphi}=\varphi$],    
$\psi=$constant, which is a SUSY cycle in the IR, we have   
\beq   
F_2\Bigg\vert_{\Sigma_2}=-2g_sN\sqrt{\alpha'}\sin\theta d\theta\wedge d\varphi .   
\eeq   
As we will see below, under the non-abelian T-duality,   
these two terms in $F_2$ will not be mapped to the same dual flux.   
We require that the flux on $\Sigma_2$ is quantised in the usual fashion
\beq
-\int F_2 = 2\kappa_{10}T_6 N.
\eeq
To achieve this we use,
\beq
T_p = \frac{1}{(2\pi)^p \alpha'\!~^{\frac{p+1}{2}}g_s},~~~~ 2\kappa_{10}= 4 (2\pi)^7 \alpha'\!~^4 g_s^2.
\eeq
So that we may associate the charge of the D6 branes $N$ with an $SU(N)$ gauge group in the dual QFT.
\subsection{G-Structures: from $G_2$ to $SU(3)$}\label{sectionstructurezz}   
We derive the G-structures and SUSY conditions at each step going from   
M-theory to type-IIA. For clarity in presentation, in this section $g_s=\alpha'=N=1$. 
   
As is shown in \cite{Cvetic:2001kp}, the M-theory background    
obeys the condition of $G_2$ holonomy.   
Hence, following \cite{Cvetic:2001kp}, but in notation suggestive    
of dimensional reduction, we introduce a set of vielbeins    
for the 7d internal space as defined in eq.(\ref{flux2a})--
here we call $z=\psi_+$,   
\beq   
\begin{array}{l l}   
\vspace{3 mm}   
&\hat{e}^r=dr,~~~ \hat{\tilde{e}}^{\theta}=b\sigma_1,
~~~\hat{\tilde{e}}^{\varphi}=b\sigma_2,~~~ \hat{e}^{z}=e^{2\phi/3}(dz+A_1)\\   
&\hat{\tilde{e}}^1=a(\Sigma_1+g\sigma_1),~~~\hat{\tilde{e}}^2=a(\Sigma_2+g\sigma_2),~~~\hat{e}^3=h(\Sigma_3-\sigma_3).   
\end{array}   
\eeq    
Here we have  used the  definitions introduced in    
previous sections (the reason for the cluttered by tildes definition will 
become clear shortly).    
The following three-form    
can be constructed from the 
projections on the SUSY spinor, needed to derive the BPS system 
\cite{Cvetic:2001kp}, 
\beq\label{eq: viel11}   
\tilde{\Phi}_3=\hat{e}^r\wedge(\hat{\tilde{e}}^{1 \theta}+\hat{\tilde{e}}^{2\varphi}   
+e^{3z})+(\hat{\tilde{e}}^{12}-\hat{\tilde{e}}^{\theta\varphi})   
\wedge(\alpha \hat{e}^3+\beta\hat{e}^z)+(\hat{\tilde{e}}^{1\varphi}-\hat{\tilde{e}}^{2\theta})   
\wedge(\alpha \hat{e}^3-\beta\hat{e}^z)   
\eeq   
where   
\beq   
\alpha(r)=\frac{a g}{\sqrt{b^2+a^2g^2}},~~~\beta(r)=\frac{b}{\sqrt{b^2+a^2g^2}},~~~\alpha^2+\beta^2=1.   
\eeq   
It is then simple to show that the three-form obeys   
\beq\label{eq: 3-form}   
d\tilde{\Phi}_3=0,~~~d\star_7\tilde{\Phi}_3=0 , 
\eeq   
once the BPS equations (\ref{eq:Deqns}) are imposed.   
We would now like to dimensionally reduce the $G_2$ SUSY conditions to find the    
corresponding conditions in type-IIA. Fortunately, this was done in full generality    
in \cite{Kaste:2002xs} and in a rather similar scenario    
in \cite{Gaillard:2009kz}. The corresponding conditions are those of an $SU(3)$-structure.    
All one needs  is to convert eq.(\ref{eq: viel11})    
to Scherk-Schwarz gauge then follow the prescription of \cite{Kaste:2002xs}.    
This is achieved through a rotation in both the   
$\hat{\tilde{e}}^{\theta},\hat{\tilde{e}}^{\varphi}$ and $\hat{\tilde{e}}^{1},\hat{\tilde{e}}^{1}$ planes such that:   
\beq   
\begin{array}{ll}   
\vspace{3 mm}   
\hat{e}^{\theta}&=\cos\psi \hat{\tilde{e}}^{\theta} - \sin\psi \hat{\tilde{e}}^{\varphi}=b d\theta\\   
\vspace{3 mm}   
\hat{e}^{\varphi}&=\sin\psi \hat{\tilde{e}}^{\theta} + \cos\psi \hat{\tilde{e}}^{\varphi}=b \sin\theta d\varphi\\   
\vspace{3 mm}   
\hat{e}^{1}&=\cos\psi \hat{\tilde{e}}^{1} - \sin\psi \hat{\tilde{e}}^{2}=a(\omega^1+g d\theta)\\   
\vspace{3 mm}   
\hat{e}^{2}&=\sin\psi \hat{\tilde{e}}^{1}    
+ \cos\psi \hat{\tilde{e}}^{2}=a(\omega^2+g\sin\theta d\varphi) .\\   
\end{array}   
\eeq    
The corresponding three-form, $\Phi_3$ is the same as eq.(\ref{eq: viel11})    
with $\hat{\tilde{e}}\to\hat{e}$ and is obviously still both closed and co-closed.   
The  vielbeins of the new 6-d internal space can be neatly expressed as   
\beq   
\begin{array}{l l}   
\vspace{3 mm}   
&e^{r}=e^{\phi/3}dr,~~~e^{\theta}=e^{\phi/3}b d\theta,~~~e^{\varphi}=e^{\phi/3}b \sin\theta d\varphi\\   
&e^{1}=e^{\phi/3}a(\tilde\omega^1+g d\theta),~~~e^{2}=e^{\phi/3}   
a(\tilde\omega^2+g\sin\theta d\varphi),~~~e^{3}=e^{\phi/3}h(\tilde\omega^3-   
\cos\theta d\varphi),   
\end{array}   
\eeq   
while the 11-D vielbeins are of the form $\hat{e}^A=(e^a,\hat{e}^z)$.    
The $SU(3)$ structure is then given in terms of the 3-form by:   
\beq   
J_{ab}= \Phi_{abz},~~~~~~(\Omega_{hol})_{abc}=\Phi_{abc}-i (\star_6\Phi)_{abc}.   
\eeq   
which amounts in this case to   
\beq   
\begin{array}{ll}   
\vspace{3 mm}   
J=&-e^{3r}+(\alpha e^2+\beta e^{\varphi})\wedge    
e^{\theta}+e^{1}\wedge(-\alpha e^{\varphi}+\beta e^{2})\\   
\Omega_{hol}=&(-e^3+i e^r)\wedge((\alpha e^2+\beta e^{\varphi})+ie^{\theta})   
\wedge(e^1+i(-\alpha e^{\varphi}+\beta e^{2})).   
\end{array}   
\eeq   
These can be used to construct two pure-spinors,   
\beq   
\Psi_+=\frac{e^{A}}{8}e^{-i J},~~~~\Psi_-=\frac{e^{A}}{8}\Omega_{hol},   
\eeq 
that can be shown to satisfy the pure spinors SUSY conditions
\beq   
\begin{array}{ll}   
\vspace{3 mm}   
&d(e^{2A-\phi}\Psi_+)=0\\   
&d(e^{2A-\phi}\Psi_-)=e^{2A-\phi} dA\wedge \bar{\Psi}_-+ i\frac{e^{3A}}{8}\star_6F_2 ,  
\end{array}   
\eeq   
which, collecting forms of equal size, gives   
\beq   
\begin{array}{ll}   
\vspace{3 mm}   
&dJ=0\\   
\vspace{3 mm}   
&d(e^{3A-\phi})=0\\   
\vspace{3 mm}   
&d(e^{2A-\phi}Re\Omega_{hol})=0\\   
\vspace{3 mm}   
&d(e^{4A-\phi}Im\Omega_{hol})-e^{4A}\star_6F_2=0   
\label{sarasa}
\end{array}   
\eeq   
these relations are all satisfied once eqs.(\ref{eq:Deqns}) 
are taken into account. We will choose $3A=\phi$. 
Also, notice that $F_4=0$ for backgrounds of $SU(3)$-structure.  
\subsubsection{Potential and Calibrations.}   
It is useful to derive an expression for  the seven form $C_7$  
that acts as a potential for $F_8$, {i.e.} $F_8=\star F_2=dC_7$. One finds,   
\beq   
C_7=e^{4A-\phi}vol_4\wedge Im\Omega_{hol}.   
\eeq   
The calibration form of space-time filling D branes is given by
\cite {Grana:2004bg},   
\beq   
\Psi_{cal}=-8e^{3A-\phi}\left(Im\Psi_-\right)=-e^{4A-\phi} Im\Omega_{hol}.   
\eeq   
Clearly we have $vol_4\wedge\Psi_{cal} +C_7=0$ so any space-time filling D6    
brane wrapping a 3-cycle $\Sigma^3$ such that the calibration condition   
\beq   
e^{4A-\phi}\sqrt{detG_{\Sigma^3}}=e^{4A-\phi} Im \Omega_{hol}\bigg\vert_{\Sigma^3}   
\label{calibrationzzzz}\eeq   
is satisfied will be SUSY\footnote{We work in conventions where the DBI and WZ actions have a relative sign difference}. The same condition must be satisfied for any    
odd cycle and so the only non vanishing odd cycles are 3-cycles    
(if $B_2$ were turned on we could also have 5-cycles).    
A similar calculation shows that potential even SUSY cycles    
are $\Sigma^2$, $\Sigma^6$ such that (these are calibrated by $Im \Psi_+$),   
\beq   
\sqrt{detG_{\Sigma^2}}=J\bigg\vert_{\Sigma^2},~~~~~~~~~   
\sqrt{detG_{\Sigma^2}}=-\frac{1}{6}J\wedge J\wedge J\bigg\vert_{\Sigma^6} .   
\eeq    
   
All the information above, only relies on the backgrounds in eqs.(\ref{d7ans0}), (\ref{metricxxx})    
and their BPS equations \eqref{eq:Deqns}.   
We will now describe some solutions to this system of first order,    
ordinary and non-linear equations.   
\subsection{Explicit Solutions}\label{explicitsolutions}   
Let us first describe a couple of known exact solutions. There is a simple   
solution to     
eqs.\eqref{eq:Deqns} given by,   
\begin{eqnarray}  
	a(r)=c(r)=-\frac{r}{3}, \;\;\;\;\;   
	b(r)= f(r)= \frac{r}{2\sqrt{3}}.   
\label{eq:exactsol}   
\end{eqnarray}   
This solution corresponds to a  $R^{1,4}\times \mathcal{M}_7$ space with metric \eqref{d7ans0},  
\be\label{eq:exact-sol-metric}   
ds_{11}^2 = dx_{1,3}^2 + dr^2 + \frac{r^2}{9}\, [(\Sigma_1-\frac{1}{2}\sigma_1)^2 +   
(\Sigma_2 -\frac{1}{2}\sigma_2)^2] + \frac{r^2}{12} (\sigma_1^2 + \sigma_2^2) +   
\frac{r^2}{9} (\Sigma_3 -\frac{1}{2} \sigma_3)^2 + \frac{r^2}{12}\sigma_3^2.   
\ee   
When  reduced to ten dimensions the    
resulting IIA dilaton behaves as   
$e^{4 \phi/3}\sim r^2$. This solution present 
a singularity at $r=0$ and the need to    
lift this background to M-theory for large values of the radial coordinate, to avoid
strong coupling in IIA.    
This solution is the `unresolved' version of the one     
written in---for example--   
eqs.(3.16)-(3.17) of \cite{Edelstein:2001pu}.   
In that case, we will have   
\bea   
dr=\frac{d\r}{\sqrt{1-\frac{a^3}{\r^3}}},\;\;\;   
b^2=f^2=\frac{\r^2}{12},\;\;\;\; a^2=c^2=\frac{\r^2}{9}(1-\frac{a^3}{\r^3}).   
\label{exactedels}   
\eea   
This solution avoids the singularity by ending the space at   
$\r=a$. Still, the behavior of the dilaton is such that   
it the Type IIA description is strongly coupled for large values of the    
radial coordinate $r$.   
To avoid this last issue and to have a background fully contained in type IIA, 
we will describe new solutions that are both non-singular   
and with bounded dilaton. These new solutions, turn out to not be known in exact form, but   
semi-analytically, that is as series expansions for large and small values of $r$,   
complemented with a careful numerical interpolation. We will study them below.   
\subsection{Semi-analytical solutions.} \label{sec:semi-analytical}
Since our goal is to work with trustable backgrounds in Type IIA   
we will be mostly interested in      
solutions with bounded dilaton and everywhere  finite Ricci and Riemann invariants.    
The  asymptotic large radius  $r\rightarrow \infty$, form of these solutions   is,   
\begin{align} \label{eq:sol-UV}   
	a(r)&=	\frac{r}{\sqrt{6}}-\frac{\sqrt{3}  {q_1} R_1}{\sqrt{2}}+\frac{21 \sqrt{3}   
		{R_1}^2}{\sqrt{2}\  16 r}+\frac{63 \sqrt{3}   {q_1}   {R_1}^3}{\sqrt{2}\ 16 r^2}+\frac{9   
		\sqrt{3} \left(672   {q_1}^2+221\right)   {R_1}^4}{\sqrt{2}\ 512 r^3}+\nonumber\\   
  & \frac{81 \sqrt{3}   {q_1} \left(224   {q_1}^2+221\right)   {R_1}^5}{\sqrt{2}\  512   
   r^4}+\frac{\sqrt{3} \left(2048    {h_1}+1377 \left(768   {q_1}^4+1632   
		   {q_1}^2+137\right)   {R_1}^6\right)}{\sqrt{2}\  8192 r^5}+ \dots\nonumber\\   
b(r) &=	\frac{r}{\sqrt{6}}-\frac{\sqrt{3}  {q_1} R_1}{\sqrt{2}}-\frac{3\sqrt{3} R_1^2}{4\ \sqrt{2} r}- \frac{9 \sqrt{3} q_1 R_1^3}{4\ \sqrt{2} r^2}-\frac{ 9 \sqrt{3}    
(37+96 q_1^2) R1^4}{128\ \sqrt{2} r^3} -\frac{81 \sqrt{3} q_1 (37+32 q_1^2) R_1^5}{128\ \sqrt{2}\ r^4}+\nonumber\\	      
&\frac{\sqrt{3} (512 h_1-81 (133+1920 q_1^2+960 q_1^4) R_1^6)}{2048 \sqrt{2}\  r^5}+\dots\nonumber\\		   
c(r) &=-\frac{r}{3}+q_1 R_1-\frac{9 R_1^2}{8 r}-\frac{27 q_1 R_1^3}{8 r^2}   
-\frac{9 (17+36 q_1^2) R_1^4}{32 r^3}-\frac{81 q_1 (17+12 q_1^2) R_1^5}{32 r^4}+\frac{h_1}{r^5}+\dots\nonumber\\   
f(r)&=  {R_1}-\frac{27  {R_1}^3}{8 r^2}-\frac{81  {q_1}   {R_1}^4}{4   
		r^3}-\frac{243 {R_1}^5\left(12  {q_1}^2  +  1\right)}{32 r^4}-\frac{729 {R_1}^6 \left(4   
     {q_1}^3  +  {q_1} \right)}{8   
   r^5}+ \dots   
\end{align}   
where $q_1, R_1$ and $h_1$ are constants.    

\begin{figure}	   
	\begin{center}   
\includegraphics[height=1.6in]{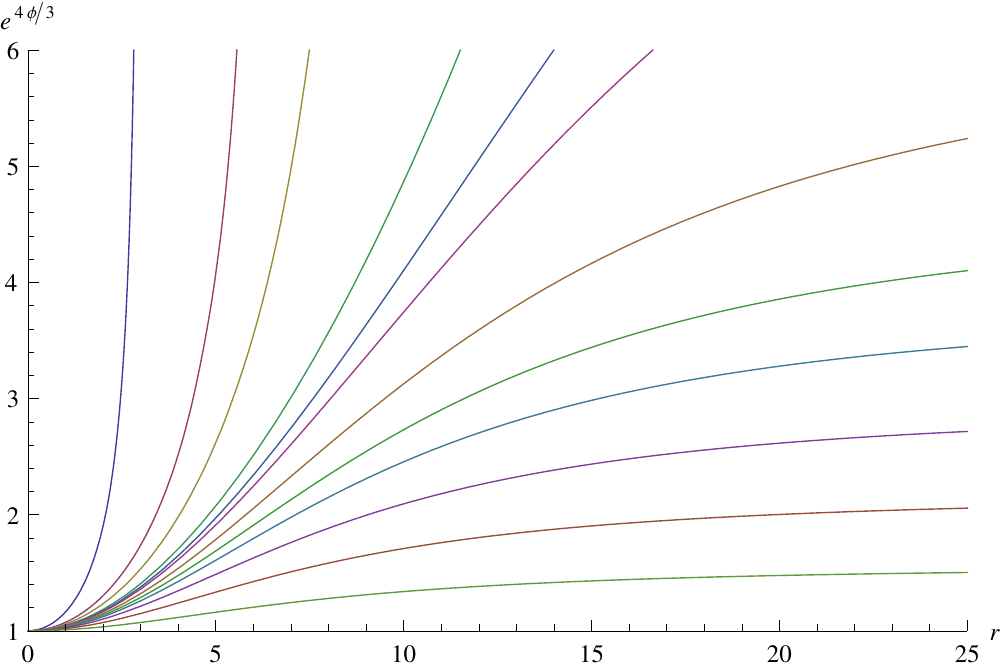}   
\caption{$e^{\frac{4 \phi }{3}}$ for different values of $q_0$  and $R_0$. We keep $q_0 R_0 =2$ fixed which amounts to fixing the normalization of the dilaton in the IR. }   
\label{fig:dilaton-R0}   
\end{center}   
\end{figure}

\begin{figure}[p] 
\begin{center} 
 \mbox{ 
\subfigure[] {\includegraphics[angle=0, 
 width=0.45\textwidth]{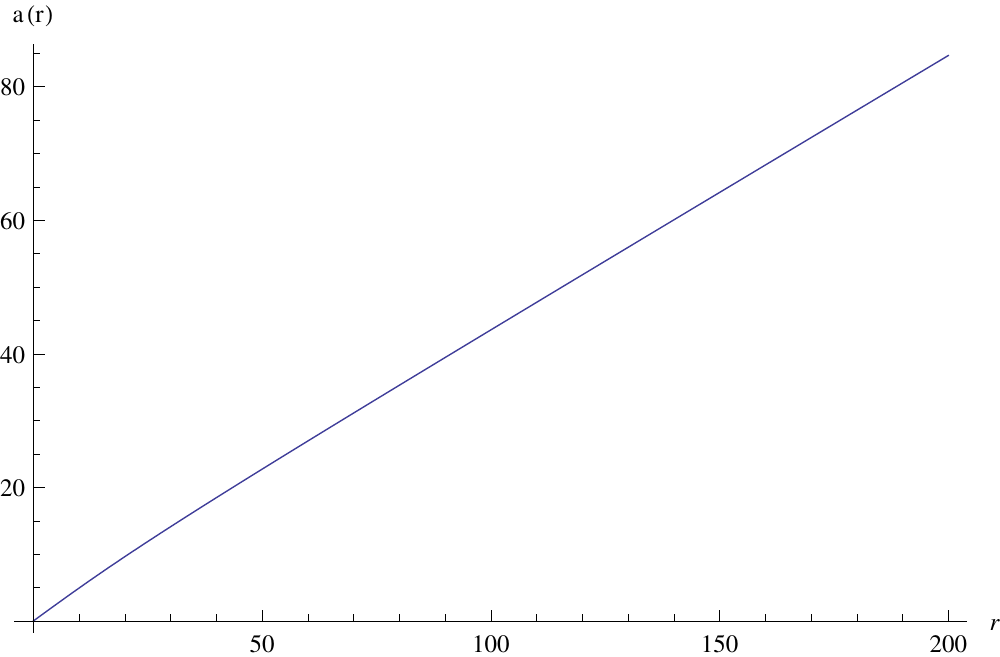} } 
 \subfigure[] {\includegraphics[angle=0, 
width=0.45\textwidth]{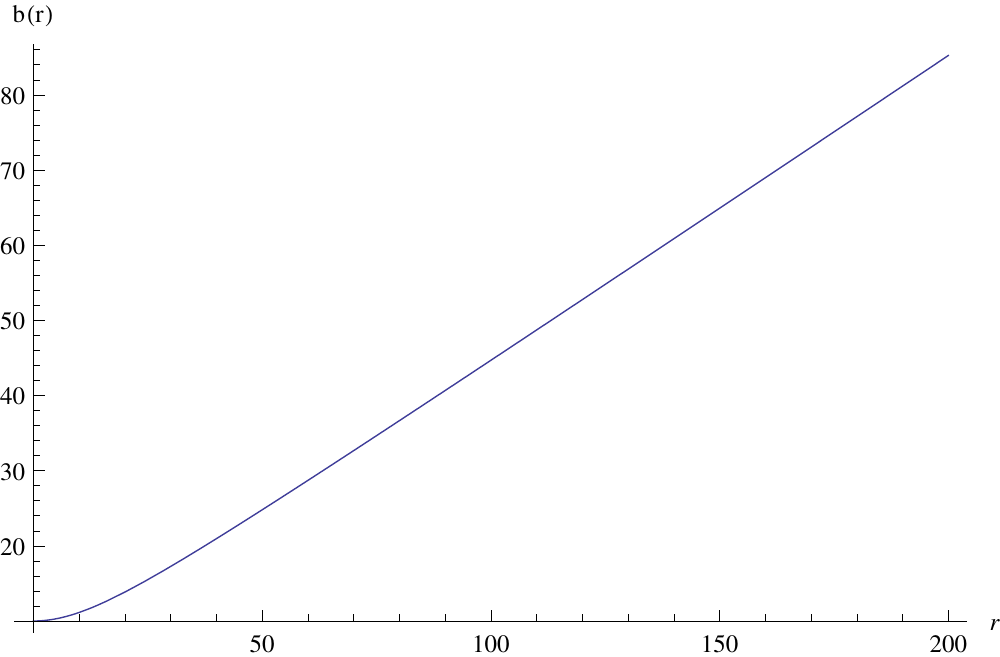} } 
} 
\mbox{ 
\subfigure[] {\includegraphics[angle=0, 
width=0.45\textwidth]{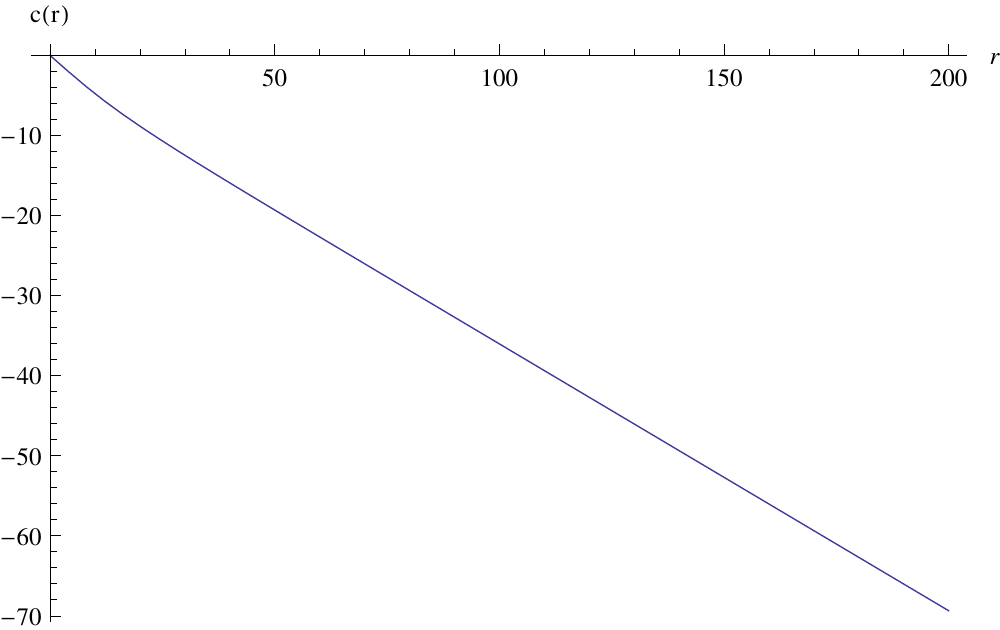} } 
 \subfigure[] {\includegraphics[angle=0, 
width=0.45\textwidth]{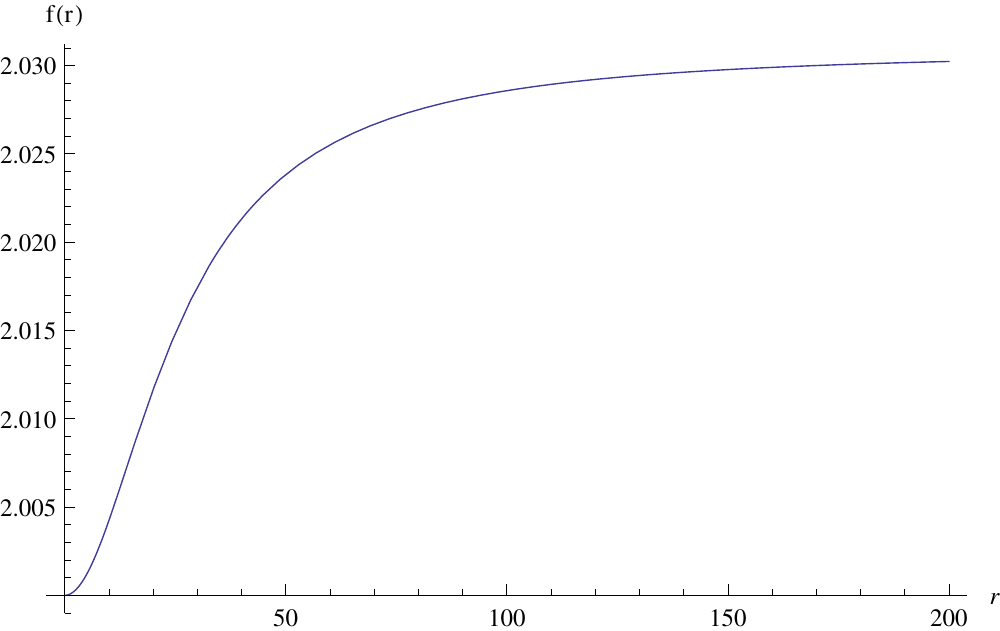} } 
} 
\mbox{ 
\subfigure[] {\includegraphics[angle=0, 
width=0.45\textwidth]{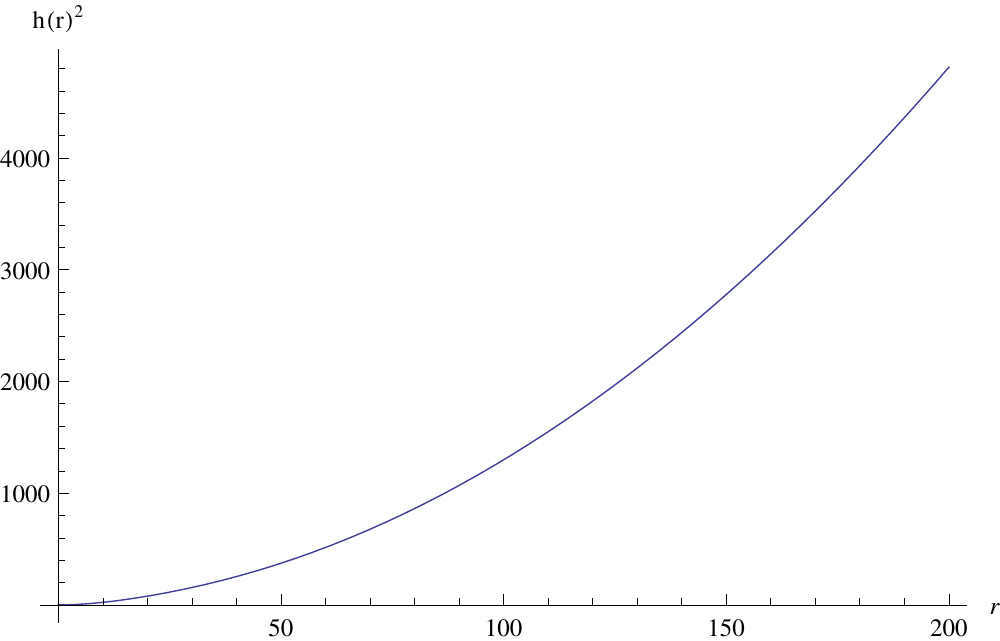} } 
 \subfigure[] {\includegraphics[angle=0, 
 width=0.45\textwidth]{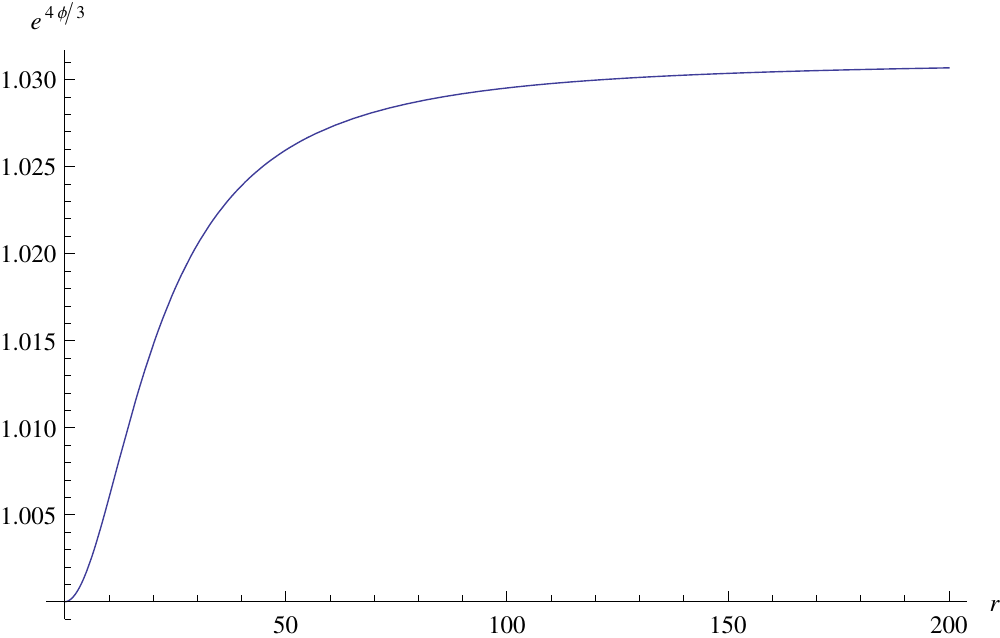} } 
}
\caption{\small A numerical solution for $a(r), b(r), c(r)$ and $f(r)$ obtained by 
forward integration of the BPS equations 
with \eqref{eq:Dat0-bis} as boundary conditions, 
$R_0=10,$ and $ q_0=1/5 $. After the minimization procedure explained in the appendix \ref{appendix-numerics} we find that for the UV parameters $q_1 =1.31946 ,\ R_1= -2.03087,\ h_1 =−1.9733$ this solution has the required UV behavior \eqref{eq:sol-UV}. We also plot $h(r)^2$ and $e^{4\phi/3}$ defined in \eqref{metricxxx}} 
 \label{fig:sols} 
\end{center} 
\end{figure}

Close to $r\to 0$ one has   
\bea\label{eq:Dat0-bis}   
a(r) & = & \frac{r}{2}-\frac{(q_0^2+2)r^3}{288 R_0^2} -   
\frac{(-74-29q_0^2+31q_0^4)r^5}{69120 R_0^4} + \cdots  , \nn \\   
b(r) & = & R_0 - \frac{(q_0^2-2)r^2}{16 R_0} -   
\frac{(13-21q_0^2+11q_0^4) r^4}{1152 R_0^3} + \cdots  , \nn \\   
c(r) & = & -\frac{r}{2} - \frac{(5q_0^2-8)r^3}{288 R_0^2} -   
\frac{(232-353q_0^2+157q_0^4) r^5}{34560 R_0^4}+ \cdots , \nn \\   
f(r) & = & q_0 R_0 + \frac{q_0^3 r^2}{16 R_0} +   
\frac{q_0^3(-14+11q_0^2) r^4}{1152 R_0^3} + \cdots ,\\   
g(r) & = & \frac{q_0}{2}+\frac{q_0(q_0^2-1)}{24R_0^2}r^2    +\cdots,\nn \\   
g_3(r) & = &  \frac{q_0^2 -2}{2} +\frac{q_0^2(q_0^2-1)}{12 R_0^2}r^2  +\cdots.\nn    
\eea   
Note that $a(r)$ and $c(r)$ collapse in the IR  and the other two functions do not.  
The constants $q_0$ and $R_0$  determine the IR behavior. Similarly,    $q_1, R_1$    
and $h_1$ are the UV parameters. Not for every set of  ${q_0, R_0, q_1, R_1, h_1}$ there will exist  a solution that interpolates    between  
\eqref{eq:sol-UV} and  \eqref{eq:Dat0-bis}. 
For example, as seen in Figure \ref{fig:dilaton-R0},    
if we numerically integrate forward from the IR, not every value of $R_0, q_0$    leads to a stabilized dilaton. Similarly if we integrate back from the UV    
using eq.\eqref{eq:sol-UV} as boundary conditions we do not necessarily  get to an IR like    
that in eq.\eqref{eq:Dat0-bis}. Nevertheless, it is possible to show numerically that solutions    
interpolating between the behavior of eqs.\eqref{eq:sol-UV} and \eqref{eq:Dat0-bis}    
do exist. 
In Figure \ref{fig:sols}   
we present representatives of such solutions. To obtain these numerical solutions    
we shoot from the IR and minimize     
the mismatch between this forward integrated solution and    
the required UV behavior. This minimization procedure  determines the  UV    
parameters (see  Appendix \ref{appendix-numerics} for more details).   
Also, we have defined some other functions in terms of the above, their expansions read, for 
$r\to \infty$
\beq
\begin{array}{ll}\label{eq:expansionsUV}
e^{4A} =&(g_sN)^{3/2}e^{4\phi/3}\\[ 2mm]
e^{4A} =&\frac{R_1^2}{4}-\frac{9 R_1^4}{8 r^2}+O\left(\frac{1}{r}\right)^3\\[ 2mm]
        h^2=&\frac{r^2}{9}-\frac{2}{3} r \left(q_1 R_1\right)+\frac{1}{2} \left(2 q_1^2+1\right)
   R_1^2+O\left(\frac{1}{r}\right)^1\\[ 2mm]
K=&\frac{6561 R_1^4}{64 r^4}+O\left(\frac{1}{r}\right)^5\\[ 2mm]
g=&\frac{3 R_1}{2 r}+\frac{9 q_1 R_1^2}{2 r^2}+\frac{27 \left(16 q_1^2-1\right) R_1^3}{32
   r^3}+O\left(\frac{1}{r}\right)^4\\[ 2mm]
g_3=&-1+\frac{9 R_1^2}{2 r^2}+\frac{27 q_1 R_1^3}{r^3}+\frac{81 \left(24 q_1^2-1\right) R_1^4}{16
   r^4}+O\left(\frac{1}{r}\right)^5\\
\end{array}
\eeq
and for $r\to 0$, we have
\beq
\begin{array}{ll}
\label{eq:expansionsIR}
e^{4A}=&\frac{1}{4} q_0^2 R_0^2+\frac{3}{64} q_0^4 r^2+\frac{q_0^4 \left(37 q_0^2-40\right) r^4}{3072
   R_0^2}+O\left(r^6\right)\\
h^2=&\frac{r^2}{4}+\frac{\left(q_0^2-16\right) r^4}{576 R_0^2}+O\left(r^6\right)\\
K=&1-\frac{r^2}{4 R_0^2}+\frac{\left(40-7 q_0^2\right) r^4}{576 R_0^4}+O\left(r^6\right)\\
g=&\frac{q_0}{2}+\frac{q_0 \left(q_0^2-1\right) r^2}{24 R_0^2}
+\frac{q_0 \left(91 q_0^4-179 q_0^2+88\right)
   r^4}{13824 R_0^4}+O\left(r^6\right)\\
g_3=&\frac{1}{2} \left(q_0^2-2\right)+\frac{q_0^2 \left(q_0^2-1\right) r^2}{12 R_0^2}
+\frac{q_0^2 \left(115 q_0^4-227
   q_0^2+112\right) r^4}{6912 R_0^4}+O\left(r^6\right)\\
\end{array}
\eeq

The numerical solutions presented in Figures \ref{fig:sols} 
satisfy $R_0 q_0=2$. This corresponds to choosing 
the normalization of the dilaton such 
that $(g_s N)^{3/2} e^{4 \phi_0/3} =1,  
$ where $\phi_0$ is the value of the dilaton at $r=0$. 
Also, since we want solutions with monotonically increasing dilaton, 
we  require  comparing  eq.\eqref{eq:expansionsUV} with 
eq.(\ref{eq:expansionsIR}), that
$R_1^2> q_0^2 R_0^2$. 

%
\subsubsection{Asymptotic behaviour}   
After reducing to ten dimensions the simple exact solution    
mentioned above   
leads to a background with metric easily obtained from  eq.(\ref{eq:exact-sol-metric}),   
 dilaton $e^\frac{4 \phi}{3} = \frac{r^2}{36 (g_sN)^{2/3} }$    
and $F_2= -\sqrt{a}'g_sN(\sin\theta d\theta\wedge d\varphi +\tilde\omega^1\wedge\tilde\omega^2)$.   
Notice that the space is not asymptotically $T^{1,1}$ for 
the exact solutions.   
On the other hand, the numerical solutions with stabilized dilaton behave in the UV as,   
\begin{align}   
	ds_{IIA,st}^2&=\alpha' g_s N \frac{R_1}{2}\Bigg[\frac{\mu    
dx_{1,3}^2}{\alpha' g_s N } + dr^2 + r^2 \Big(\frac{1}{6}(d\theta^2 + \sin^2\theta d\varphi^2   
	+(\tilde\omega^1)^2+(\tilde\omega^2)^2)\nn\\   
	&\hspace{3.5in}+ \frac{1}{9}(\tilde\omega^3-\cos\theta d\varphi)^2\Big) + \dots]   
\end{align}   
with    
\begin{align}   
	(g_s N)^{2/3}e^\frac{4 \phi}{3} &= \frac{R_1^2}{4}-\frac{9 R_1^4 }{8 r^2} -\frac{27 q_1 R_1^5}{4 r^3} \dots\nn\\   
	F_2&= -\sqrt{\a}'g_sN[(1- \frac{81 R_1^2}{8 r^2}+\dots)\sin\theta d\theta\wedge    
d\varphi + (1 + \frac{81 R_1^2}{8 r^2}+\dots)\tilde\omega^1\wedge\tilde\omega^2 \nn\\   
	&\hspace{2.5in}-(\frac{81 R_1^2}{4 r^3}+\dots) dr\wedge(\tilde\omega^3-\cos\theta d\varphi)]\nn\\   
	&\sim-\sqrt{\a '}g_sN(\sin\theta d\theta\wedge d\varphi + \tilde\omega^1\wedge\tilde\omega^2 )   
\end{align}   
In the UV the five dimensional internal space is $T^{1,1}$.    
Thus, the space is asymptotically $R^{1,3}\times CY_6$ with    
a constant dilaton and constant $F_2$.   
This `flat space' asymptotics is    
characteristic of duals to QFTs whose UV behavior   
is controlled by an irrelevant operator---this will    
come back when dealing with the QFT analysis.    
Somehow the field theory is taken out of the `decoupling limit'.    
On the other hand, in the IR the metric, dilaton and RR form asymptote to,   
\beq   
\begin{array}{ll}   
\vspace{3 mm}   
ds^2_{IIA,str}&=\frac{q_0 R_0 \alpha' g_s N}{2}\bigg[\mu dx^2_{1,3}+ dr^2+R_0^2\big(d\theta^2+\sin^2\theta d\varphi^2\big) + \frac{r^2}{4}d\Omega_3 \bigg]+...\\   
\vspace{3 mm}   
d\Omega_3 &= (\tilde\omega_1+d\theta)^2+(\tilde\omega_2+ \sin\theta d\varphi)^2 +(\tilde\omega_3 -\cos\theta d\varphi)^2,    
\;\;\;(g_s N)^{2/3}e^{4\phi/3}=\frac{q_0^2 R_0^2}{4}+\frac{3q_0^4}{64}r^2+...\\   
F_2&=-2 \sqrt{\alpha}' g_s N \sin\theta d\theta\wedge d\varphi+...   
\end{array}   
\eeq

The material  discussed in this section is not all original; 
we have rewritten some of it to ease the analysis of the next section.  
However, we should point out that the semi-analytic solutions with stabilized 
dilaton and no singularities (though have been discussed in \cite{Brandhuber:2001kq}   
and \cite{Cvetic:2001kp}) are found explicitly---
with the explicit delicate numerics---in this paper. 
These solutions will play an important role in the next sections. 
   
\section{Non-Abelian T-duality.} \label{NATDx}   
   
In this section, we will present completely original   
material. We will construct a {\it new } solution   
in Type IIB supergravity preserving four supercharges. 
This background will have    
$SU(2)$-dynamical structure. We believe this type of solution  is    
new in the literature.    
   
The technique we will use to construct this new background is non-Abelian T-duality, see   
\cite{de la Ossa:1992vc}-\cite{Lozano:2012au} for a partial sample of papers.   
The reader unfamiliar with this technology should read    
Section 2 in \cite{Itsios:2013wd} for a clear explanation of the whole procedure.   
   
We will straightforwardly present the new background in type IIB supergravity.   
Following the conventions of    
Section 2 of the paper 
\cite{Itsios:2013wd} and starting from the background in   
eq.(\ref{metricxxx}) we perform    
a non-Abelian T-duality on the $SU(2)$ isometry parametrised by    
$(\theta,\varphi,\psi)$ and gauge fix such 
that $\theta=\varphi=v_1=0$, so that the solution generated 
still depends on the angles $(\tilde{\theta},\tilde{\varphi}, \psi)$
and on the new coordinates $(v_2, v_3)$---see the short
discussion below eq.(\ref{IIaflux}) about the explicit $SU(2)$
invariances of the background. We remind the reader that 
$\tilde{\omega}^i$,
\beq
\tilde{\omega}_1= \cos \psi\, d \tilde{\theta} 
+ \sin \psi\, \sin \tilde{\theta}\, d\tilde{\varphi},
~~\tilde{\omega}_2= -\sin \psi\, d \tilde{\theta}    
+ \cos \psi\, \sin \tilde{\theta}\, d\tilde{\varphi},~~\tilde{\omega}_3= 
d\psi + \cos \tilde{\theta}d\tilde{\varphi}.
\eeq    
In the process of doing    
this non-Abelian T-duality, we generate an entirely new NS and RR sector and type-IIB metric.   
The T-dual metric is given by (we take $g_s=\alpha'=\mu=1$ and we 
remind the reader that below eq.(\ref{sarasa}) we set $3A=\phi$),   
\beq   
\begin{array}{ll}   
\vspace{3 mm}   
ds_{IIB,st}^2&\!=e^{2A}\Bigg[dx_{1,3}^2 + N dr^2 + 
N \hat{a}^2(d\tilde{\theta}^2 + 
\sin^2\tilde{\theta} d\tilde{\varphi}^2)\Bigg]+   
\frac{1}{\det{M}}\Bigg[2 (v_3 dv_2\!+\!v_3 dv_3)^2\!+\\   
&~~~4N^2e^{4A} \hat{b}\bigg(\hat{b}^2(dv_3+\hat{c}v_2\tilde{\o}_2)^2+
h^2(\hat{c}^2v_3^2 \tilde{\o}_1^2+ (dv_2-\hat{c}v_3\tilde{\o}_2)^2)
+2\hat{c}v_2v_3\tilde{\o}_1\tilde{\o}_3+v_2^2\tilde{\o}_3)\bigg)\Bigg]
\label{iibmetricxxx}\end{array}   
\eeq   
where   
\beq   
\det{M}=4 e^{2 A} N\left(2e^{4A}N^2 \hat{b}^4h^2+\hat{b}^2v_2^2+h^2v_3^2\right),   
\eeq   
which also appears in the definition of the dual dilaton 
\beq   
e^{-2\Phi}=\det{M} e^{-2\phi},   
\eeq   
and we have introduced the following functions for convenience of presentation
\beq   
\hat{a}=\frac{ab}{\sqrt{b^2+a^2g^2}},~~~\hat{b}=\sqrt{b^2+a^2g^2},~~~\hat{c}=\frac{a^2g}{b^2+a^2g^2}.
\eeq   
The many and  
complicated forms that this background supports 
can be expressed in a relatively compact manner 
through a judicious choice of dual vielbein basis $\hat{e}^a$, namely
\beq
\begin{array}{l l}\label{eq: duale}
&e^{x^\mu}= e^{A} dx^{\mu},~~~~~
e^{r}= e^{A}\sqrt{N} dr,~~~~~ 
e^{1,2}=e^{A}\sqrt{N} \hat{a}\tilde{\omega}_{1,2}\\[3 mm]
&e^{\hat{1}}=\frac{2\sqrt{N}e^{A}\hat{b}}{\det M}\bigg[-\sqrt{2}v_2(v_2dv_2+v_3dv_3)-2\sqrt{2} e^{4A}N^2\hat{b}^2h^2(dv_2-\hat{c}v_3\tilde{\o}_2)+\\
&~~~~~~~~~~~~~~~~~~~~~~~~~~~~~~~~~~~~~~~~2e^{2A}N h^2v_3(v_3\hat{c}\tilde{\o}_1+v_2 \tilde{\omega}_3)\bigg]\\[3 mm]
&e^{\hat{2}}=\frac{4e^{3A} N^{3/2}\hat{b}}{\det M}\bigg[v_2\hat{b}^2(dv_3+\hat{c}v_2\tilde{\o}_2)+h^2(\hat{c}v_3^2\tilde{\o}_2-v_3dv_2)-\\
&~~~~~~~~~~~~~~~~~~~~~~~~~~~~~~~~~~~~2\sqrt{2}e^{2A}N h^2\hat{b}^2(\hat{c}v_3\tilde{\o}_1+v_2\tilde{\o}_3)\bigg]\\[3 mm]
&e^{\hat{3}}=\frac{2e^{A}\sqrt{N}h}{\det M}\bigg[-\sqrt{2}v_2(v_2dv_2+v_3dv_3)-2\sqrt{2}e^{4A}N^2\hat{b}^4(dv_3+\hat{c}v_2\tilde{\omega}_2)-\\
&~~~~~~~~~~~~~~~~~~~~~~~~~~~~~~~~~~~~~~~~2e^{2A}N\hat{b}^2 v_2(\hat{c}v_3\tilde{\o}_1+v_2\tilde{\o}_3)\bigg].
\end{array}
\eeq
With respect to this basis the NS two-form is given by \footnote{Note that the procedure of \cite{Itsios:2013wd} actually gives the NS two from up to an exact $
B_{2,eq.(\ref{eqb2no})}=B_{2,NATD}+ \frac{1}{\sqrt{2}}d\psi \wedge dv_3$. The choice we make is merely more simple in vielbein basis.}
\beq
B_2= \frac{1}{\hat{a}\hat{b}v_2}
\bigg[\hat{a}hv_2 e^{\hat{1}\hat{3}}-\hat{b}
\hat{c}hv_3e^{1\hat{3}}-\hat{b}^2\big(\hat{c}v_2e^{1\hat{1}}
+\sqrt{2}e^{2A}N \hat{a} h e^{\hat{2}\hat{3}}\big)\bigg]
\label{eqb2no}\eeq   

The RR sector is given by,
\beq   
\begin{array}{ll}   
 \vspace{3 mm}   
F_1&= \frac{2e^{-A}\sqrt{N}(K+1)}{\hat{a}\hat{b}}\bigg[\hat{b}\big(\hat{c}v_2e^{2}+\sqrt{2}e^{2A}N \hat{a}h e^{\hat{3}}\big)-\hat{a}v_2 e^{\hat{2}}\bigg]- 2e^{-A}\sqrt{N}K' v_3 e^{r},\\[3 mm]   
F_3&=\frac{2(K+1)e^{-A}\sqrt{N}}{\hat{a}\hat{b}^2}\bigg[\hat{b}^2\hat{c}v_2 e^{1\hat{1}\hat{2}}+\hat{b}\hat{c}h v_3\big(e^{2\hat{1}\hat{3}}-e^{1\hat{2}\hat{3}}\big)-\sqrt{2}e^{2A}N \hat{b}^3 \hat{c}\big(e^{1\hat{1}\hat{3}}+e^{2\hat{2}\hat{3}}\big)+\hat{a}h v_3 e^{\hat{1}\hat{2}\hat{3}}\bigg]+\\[3 mm]
&~~~~~~\frac{2e^{-A}\sqrt{N}\hat{b}K'}{h}\bigg[\sqrt{2}e^{2A}N\hat{b}h e^{r \hat{1}\hat{2}}+v_2 e^{r \hat{1}\hat{3}}\bigg]+\frac{2e^{-A}\sqrt{N}U}{\hat{a}}\bigg[\hat{b}v_2 e^{12\hat{1}}+hv_3 e^{12\hat{3}}\bigg],\\[5 mm]
F_5&= \frac{2\sqrt{2} e^{A}N^{3/2} \hat{b} h U}{\hat{a}^2}\bigg[e^{tx^1x^2x^3}-e^{12\hat{1}\hat{2}\hat{3}}\bigg],
 \label{fluxesiibzz}\end{array}   
\eeq   
where 
\beq
U= \hat{c}(K+1)- (K-1),
\label{xxyy}
\eeq
has been defined for convenience. We also note that the potential such that $F_1=dC_0$ is actually very simple, namely $C_0=-2N(K+1)v_3$.
We have checked using Mathematica that this background solves   
the Einstein, dilaton, Maxwell and Bianchi equations of Type IIB, once   
the eqs.(\ref{eq:Deqns}) are imposed.

Notice, that like in the paper \cite{Petrini:2009ur}, 
our background's warp factors and dilaton
depend on more than one coordinate-- ($r, v_2, v_3$)-- in our case.

\subsection{Asymptotics}
In the IR the new 3 manifold that is generated has induced metric
\beq
ds_{3}^2= \frac{1}{2N q_0 R_0^3 v_2^2}\bigg(2v_2^2dv_2^2
+4v_2v_3 dv_2dv_3+(N^2q_0^2R_0^6+2v_3)dv_3^2\bigg)+...
\eeq
The form of this metric suggests that $v_2=0$ produces a singularity and indeed calculating the curvature invariants in the IR are all inversely proportional to some power of $v_2$. For instance
\beq
R= \frac{q_0^2(2N^2R_0^6-15 v_2)+4v_3^2}{2Nq_0 R_0^3 v_2^2}+...
\label{nonsingularitycondition}\eeq
One may want to restrict the range of the coordinate $v_2>0$ 
to ensure our solution is non singular. This is a physical requisite on
a coordinate, that the process of non-Abelian duality 
gives no information on. But, imposing that $v_2>0$ may lead to
a space that is not
consistent geometrically, namely the manifold would not be
well defined (probably geodesically incomplete).
It should be interesting to determine if there is any geometrical 
obstruction to such restriction. We will elaborate more on this point
below.

The appearence of this possible-singular behavior at
$v_2=0$ is due to the fact that we are T-dualising
on a manifold ($\theta, \varphi, \psi$) with a shrinking fiber $\psi$.
See eq.(\ref{metricxxx}) together with eq.(\ref{eq:expansionsIR}).
Since the non-Abelian T-duality (at least at the supergravity
level as we are doing it) does not restrict
the range of the coordinates, we may propose to restrict $v_2>0$. 
Recent developments on
the sigma model side of the formalism 
\cite{Sfetsos:2013wia}
may illuminate
these issues, but still more work on the topic is needed. 
It may be that the restriction $v_2>0$ is not feasible as discussed above
and/or generates a manifold with a boundary. In that case,
our solution would present a singularity at $v_2=0$.
Physical observables would be trustable as long as they do not
'sit' on the point $v_2=0$.

In the UV the 3 manifold  has induced metric
\beq
d^2s_{3}=\frac{3}{N R_1r^2}\bigg( 2dv_2^2+3 dv_3^2 +2 v_2(d\psi+ \cos\psi d \tilde{\theta})^2\bigg)+...
\eeq
Although this is vanishing, in line with our expectations from dualising a 
manifold which blows up, all the curvature invariants remain finite. 
Related to this is the fact that, whilst the 
induced metric $g_3$ vanishes, 
the string volume $e^{-\Phi}\sqrt{\det g_3}$ is finite.

Finally, let us quote the asymptotics of the dilaton of
Type IIB. For small values of $r$, we have
\beq
e^{\Phi}=\frac{q_0}{4 N v_2}-\frac{r^2 \left(q_0 \left(N^2 q_0^2 R_0^6+2 
\left(v_3^2-\left(q_0^2-1\right) v_2^2\right)\right)\right)}{64 \left(N R_0^2 v_2^3\right)}+....
\eeq
while the dual dilaton for $r\to \infty$ is,
\beq
e^{\Phi}=\frac{9}{\sqrt{2} N^2 r^3}+\frac{81 q_1 R_1}{\sqrt{2} N^2 r^4}+\frac{243 \sqrt{2} 
q_1^2 R_1^2}{N^2r^5}+....
\eeq

\subsection{G-structure.}   
The seed type-IIA solution of section \ref{section: IIA} 
exhibits confinement and supports an $SU(3)$ structure as discussed in 
Section \ref{sectionstructurezz}. 
The results of \cite{Gaillard:2013vsa} suggest that the T-dual solution should support a 
dynamical $SU(2)$-structure, defined by a point dependent rotation between the two 
6-d internal killing spinors. This is indeed the case, we will present the structure 
here and refer the reader to Appendix D of \cite{Gaillard:2013vsa} 
for the details of the calculation\footnote{Actually it is the isometry defined by 
($\tilde{\theta},\tilde{\varphi},\psi)$ that is dualised in  Appendix D of \cite{Gaillard:2013vsa}, 
but this calculation is completely 
analogous to our's. Our result is  non-singular in the radial coordinate $r$.}
To express the structure succinctly 
it is useful to introduce a new set of vielbeins, which 
are a rotation of eq.(\ref{eq: duale}),
\beq
	\begin{aligned}
		&e^{r}=e^{A}\sqrt{N}dr\,,~~~~~\check e^{\theta}=e^{A}\sqrt{N}\hat{a} \, d\tilde{\theta}\,,~~~~~\b  e^{\varphi} + \a  e^2 = e^{A}\sqrt{N}\hat{a} \sin\tilde{\theta} d\tilde{\varphi} \,,\\
		& e^{1'}=\frac{2e^A \sqrt{N}\hat{b}}{\det M}\bigg[-2 \sqrt{2}e^{4A}N^2\hat{b}^2 h^2 \big(\cos\psi dv_2-\hat{c}v_3(\sin\psi \tilde{\o}_1+\cos\psi \tilde{\o}_2)-v_2\sin\psi \tilde{\o}_3\big) \\
		&\qquad\qquad\quad\qquad-\sqrt{2}v_2\cos\psi(v_2 dv_2+v_3 dv_3)+2e^{2A}N \bigg(-\hat{b}^2v_2\sin\psi(dv_3+\hat{c}v_2\tilde{\o}_) \\
		&\qquad\qquad\quad\qquad\qquad\quad+ h^2v_3\big(\sin\psi dv_2+\hat{c}v_3(\cos\psi\tilde{\o}_1-\sin\psi\tilde{\o}_2)+v_2\cos\psi\tilde{\o}_3\big)\bigg)\bigg]\,\\
		&\a  e^{\varphi} - \b  e^{2'}=\frac{2e^A \sqrt{N}\hat{b}}{\det M}\bigg[-2 \sqrt{2}e^{4A}N^2\hat{b}^2 h^2 \big(\cos\psi dv_2-\hat{c}v_3(\sin\psi \tilde{\o}_1+\cos\psi \tilde{\o}_2)-v_2\sin\psi \tilde{\o}_3\big) \\
		&\qquad\qquad\qquad\qquad\quad\qquad-\sqrt{2}v_2\cos\psi(v_2 dv_2+v_3 dv_3)+2e^{2A}N \bigg(-\hat{b}^2v_2\sin\psi(dv_3+\hat{c}v_2\tilde{\o}_) \\
		&\qquad\qquad\qquad\qquad\quad\qquad\qquad\quad+ h^2v_3\big(\sin\psi dv_2+\hat{c}v_3(\cos\psi\tilde{\o}_1-\sin\psi\tilde{\o}_2)+v_2\cos\psi\tilde{\o}_3\big)\bigg)\bigg]\,\\
		& e^{3'}=\frac{2e^{A}\sqrt{N}h}{\det M}\bigg[-\sqrt{2}v_2(v_2dv_2+v_3dv_3)-2\sqrt{2}e^{4A}N^2\hat{b}^4(dv_3+\hat{c}v_2\tilde{\omega}_2)\\
		&\qquad\qquad\qquad\qquad\qquad\qquad\qquad\qquad\qquad\qquad\qquad-2e^{2A}N\hat{b}^2 v_2(\hat{c}v_3\tilde{\o}_1+v_2\tilde{\o}_3)\bigg] \,.  
	\end{aligned}
\eeq  
One then takes these vielbeins ordered as $(r\theta\varphi 1'2'3')$ 
and rotates to define another basis of vielbeins as
\beq
	\tilde e = R.e'.
\eeq
The matrix with which this rotation is performed is
\beq
	R = \frac{1}{\sqrt{\Delta}}\left(
	\begin{array}{cccccc}
		\b & 0 & 0 & \zeta_1 & -\zeta_2\b & \zeta_3 \\
		0 & \sqrt{\Delta} & 0 & 0 & 0 & 0 \\
		0 & 0 & \sqrt{\Delta} & 0 & 0 & 0 \\
		-\zeta_1 & 0 & 0 &\b & \zeta_3 & \zeta_2\b \\
		\zeta_2\b & 0 & 0 & -\zeta_3 & \b & \zeta_1 \\
		-\zeta_3 & 0 & 0 & -\zeta^2\b & -\zeta^1 & \b \\
	\end{array}
	\right)
\eeq
where
\beq
	\Delta = \b^2 +\zeta_1^2+\zeta_2^2 \b^2+\zeta^2_3
\eeq
and
\beq
\zeta_1=-\frac{e^{-2A}v_2\cos\psi}{\sqrt{2}N \hat{b} h},~~~~\zeta_2=-\frac{e^{-2A}v_2\sin\psi}{\sqrt{2}N \hat{b} h},~~~~\zeta_3=-\frac{e^{-2A}v_3}{\sqrt{2} N\hat{b}^2}.
\eeq

\begin{figure}\label{fig:alpha}
	\begin{center}
	\includegraphics[height=2in]{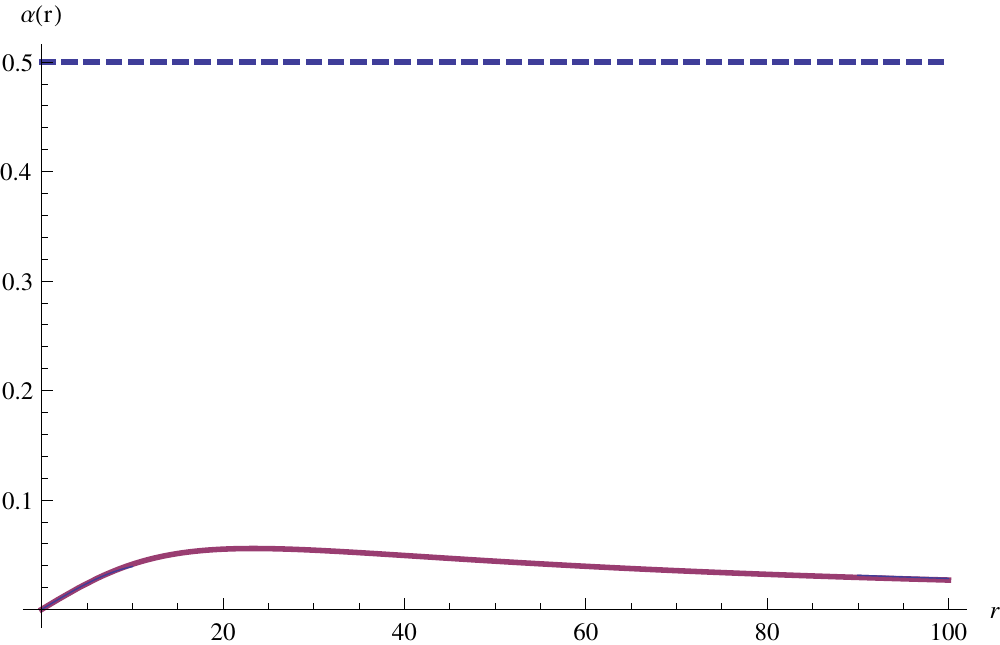}
	\caption{Solid line: $\alpha(r)$ 
for the numerical solution with $R_0=10$, $q_0= 1/5$. 
Dashed line: $\alpha(r)$ for the exact solution of 
eq.\eqref{eq:exactsol},  $\alpha_{exact}(r)=\frac{1}{2}$}
\end{center}
\end{figure}

Let us now express the forms of the geometric structure, 
following the conventions of \cite{Andriot:2008va} we have
\beq
	\begin{aligned}
		&k_{\|} = \frac{\alpha}{\sqrt{1+\zeta.\zeta}} \qquad k_{\perp} = \sqrt{\frac{\b^2 + \zeta.\zeta}{1+\zeta.\zeta}} \\
		&z = w - i \, v = \frac{1}{\sqrt{\b^2 + \zeta.\zeta}} \big( \sqrt{\Delta} \tilde e^r + \zeta_2 \alpha \tilde e^{\varphi}- i (\sqrt{\Delta} \tilde e^3 + \zeta_2 \alpha \tilde e^\theta) \big) \\
		&j = \tilde e^{r3} + \tilde e^{\varphi \theta} + \tilde e^{21} - v \wedge w \\
		&\omega =  \frac{-i}{\sqrt{\b^2 + \zeta.\zeta}} \big( \sqrt{\Delta} (\tilde e^\varphi+ i \tilde e^\theta) - \zeta_2 \alpha (\tilde e^r + i \tilde e^3) \big) \wedge (\tilde e^2 + i \tilde e^1).
	\end{aligned}
\label{useful}\eeq
In terms of those forms, we can define two 6-d pure spinors as:
\beq
	\begin{aligned}
		\Phi_+ &= \frac{i e^{A}}{8}  e^{-i v\wedge w} \big( k_{\|} e^{-ij} - i k_{\perp} \omega \big) \\
		\Phi_- &= \frac{i e^{A} }{8}  (v+i w)\wedge \big( k_{\perp} e^{-ij} + i k_{\|} \omega \big)
	\end{aligned}
\eeq
Notice that because $k_{\|}$ is point dependent we have a dynamical $SU(2)$-structure.
To have a good idea of the dynamical character
of the $SU(2)$-structure, we can expand the quantities $k_\|, k_\perp$. For the solution
 in eq.(\ref{exactedels}), we have for $\rho\to\infty$,
\beq
k_{\perp}=\frac{\sqrt{3}}{2}+\frac{\sqrt{3} a^3}{16 \rho ^3}+....,~~~~~~
k_{\|}=-\frac{1}{2}+\frac{3 a^3}{16 \rho ^3}+....
\eeq
While for $\rho\to a$ we have,
\beq
k_{\perp}=1-\frac{\left(a^4 N^2\right) (\rho -a)^2}{1728 v_2^2}+....,~~~~~~
k_{\|}=-\frac{\left(a^2 \text{Nc}\right) (\rho -a)}{12 \left(\sqrt{6} v_2\right)}+....
\eeq
On the other hand for the semi-analytic solutions we have,
\beq
k_{\perp}=1-\frac{9 R_1^2}{8 r^2}+...,~~~~~~k_{\|}=-\frac{3 R_1}{2 r}+...
\eeq
for the large radius expansion and 
\beq
k_{\perp}=1-\frac{r^4 \left(q_0^4 R_0^2\right)}{256 v_2^2}+...,~~~~~~k_{\|}=-\frac{r^2 \left(q_0^2 R_0\right)}{8 \left(\sqrt{2} v_2\right)}+...
\eeq
for the case of $r\to 0$. These expansions make clear the dynamical character of the structure.
Also very descriptive is the quantity $\alpha(r)$ shown in 
Figure \ref{fig:alpha}.

It is interesting to notice that for the non-Abelian T-dual of
the exact and singular solution in eq.(\ref{eq:exactsol}), the $SU(2)$-structure
is not dynamical. It is precisely the deformation of the space, displayed
by the non-singular solution or the 
semi-analytical ones that makes the structure dynamical. 
This may be related with the phenomena of
'confinement' and 'symmetry breaking' that occur in the dual field theory.

The calibration forms of SUSY cycles 
in the 6-d internal space are defined by
\beq
\Psi_{n}= -8 e^{-\Phi}\text{Im}
\bigg(\Phi_{\pm}\bigg)\wedge e^{-B_2}\bigg|_{n}
\eeq
where on the left hand side it should be understood that we restrict to the part with $n$-legs and the even/odd calibrations are given by $\Psi_{\pm}$ respectively. In the bibliography, one can find {\it compactifications}
with $SU(2)$-dynamical structure \cite{andriotxx}. Here we have 
constructed a non-compact manifold with that characteristic.

\subsection{SUSY sub-manifolds}\label{sec:SUSYcycles}
Here we present a list of supersymmetric sub-manifolds,
that while not exhaustive, gives at least some indication of the types of SUSY
subspaces that this type IIB solution supports.
Attention shall be restricted to manifolds with no legs in the $r$-direction.

In following sections, we will analyse different quantities derived from our
background that can be put in correspondence with observables in the dual QFT.
This analysis will {\it suggest} to impose certain conditions
on the coordinates $v_2, v_3$. Indeed, we will define sub-manifolds--that will
be referred to as '{\it cycles}', though technically they may present
boundaries. The issues of the periodicities (or not) of the coordinates
$v_2, v_3$, the presence of boundaries in our sub-manifolds, etc
are difficult to sort out in the present system and with the 
present choice of coordinates. 

Hence, the analysis in the sections below is to be read
as a field theoretical-way to get some hint on the ranges and 
periodicities (if any) of
these coordinates introduced by the dualisation procedure. A more dedicated
analysis--perhaps in a  more symmetric system \cite{nuevos}-- is in order, 
but beyond the scope of the present work.

\subsubsection*{One-cycles}
These may
be defined by imposing $v_2(v_3)$
with all other coordinates constant. The DBI action is given by
\beq
S_{DBI}^{1}= T_1\int dv_3\mathcal{L}_{DBI}^1
=T_1\int dv_3 e^{-\phi}\sqrt{2e^{4A}N^2(\hat{b}^4+ \hat{b}^2 h^2v_2' )
+(v_3 +v_2 v_2')^2}
\eeq
and the behaviour of the integrand in the IR and and UV is
\beq
\mathcal{L}_{DBI}^1= \left\{
\begin{array}{ll}
4\sqrt{\frac{N(n^2q_0^2R_0^6+2(v_3+v_2v_2')^2)}{2q_0^3 R_0^3}}+...  &~~~~\text{as}~~ r\to0\\[3 mm]
\frac{1}{3}\sqrt{\frac{2}{3}}\sqrt{\frac{N^3(3+2v_2')}{R_1}}r^2+... &~~~~\text{as}~~ r\to\infty
\end{array}\right.
\label{1-cycleiib}
\eeq
A one-cycle is SUSY when $\mathcal{L}_{DBI}^1dv_3= \Psi_1$
on that cycle. This may be used to fix $v_2(v_3)$.
The calibration 1-form on $\{v_3|v_2=v_2(v_3)\}$ is given by
\beq
\Psi_1=\frac{4 e^{6A-\phi} Nb
\hat{b}^2}{4b^2+\frac{a^4f^2}{b^2c^2}} dv_3=\left\{
\begin{array}{ll}
\frac{R_0^2(Nq_0R_0)^{3/2}}{\sqrt{2}}dv_3 +...  &~~~~\text{as}~~ r\to0\\[3 mm]
\frac{(N R_1)^{3/2}}{6\sqrt{2}}r^2dv_3+... &~~~~\text{as}~~ r\to\infty
\end{array}\right.
\eeq
It is a simple matter to show that a 1-cycle which is SUSY
in the UV is given by
\beq
v_2=\frac{1}{4}\sqrt{\frac{3}{2}}\sqrt{R_1^4-16}v_3+ C
\eeq
where $C$ is any real constant and
a real solution requires $|R_1|\geq2$
which is consistent with the numerical
solutions presented in Section \ref{sec:semi-analytical}.
Whilst there is a one cycle which is SUSY in the IR whenever
\beq
v_2^2=\frac{1}{4}\bigg(N q_0 R_0^3\sqrt{2R_0^4q_0^4-32}v_3-4v_3^2+4 C^2\bigg)
\eeq
where $C$ is a different real constant.
Notice that this simplifies to $v_2^2+v_3^2=C^2$
when $R_0q_0=2$ and then the cycle defines a circle, a similar cycle was
defined for a flavour D6 brane in \cite{Barranco:2013fza}.
\subsubsection*{Two-cycles}
There are some cycles which preserve SUSY for large values of $r$.
One of them is given by $(\tilde{\theta},v_3)$ such that
$\psi=0$ and $v_2= \frac{2\sqrt{6}}{R_1^4-16}v_3$
\footnote{Or equivalently  $(\tilde{\varphi},v_3)$
such that $\tilde{\theta}=\psi=\pi/2$,
$v_2= \frac{2\sqrt{6}}{R_1^4-16}v_3$}.
For this cycle the DBI action is obtained by integrating
\beq\label{eq:2cy1}
e^{-\Phi}\sqrt{g+B_2}\bigg|_{\Sigma_2}= B\sqrt{v_3^2+ C}
\eeq
where
\beq
\begin{array}{ll}
B&=e^{A-\phi}\frac{8+R_1^4}{R_1^4-16}
\sqrt{2N (\hat{a}^2+\hat{b}^2\hat{c}^2)}\\ [3 mm]
C&= \frac{2 e^{4A}N^2(R_1^4-16)
\hat{b}^2(24\hat{b}^2\hat{c}^2h^2+
\hat{a}^2((R_1^4-16)\hat{b}^2+24h^2))}{(r_1^4+8)^2
(\hat{a}^2+\hat{b}^2\hat{c}^2)}\\
\end{array}
\end{equation}
One can integrate this to get the volume of the cycle to behave as
\beq
\int dv_3e^{-\Phi}\sqrt{g+B_2}\bigg|_{\Sigma_2}=\left\{\begin{array}{ll}
\frac{N^2  R_1^2 \Delta v_3}{3 \sqrt{6} \sqrt{R_1^4-16}}r^3+...,&~~~~ r\to\infty\\[5 mm]
\big(\mathcal{F}(v_{3a})-\mathcal{F}(v_{3b})\big)r+...,&~~~~ r\to 0
\end{array}\right.
\eeq
where $v_{3a}, v_{3b}$ are the two values determining the 
range of the coordinate $v_3$.
\beq
\mathcal{F}(v_{3})=\frac{N  \left(R_1^4+8\right) \left(v_3 \sqrt{v_3^2+\alpha }+\alpha  \log
   \left(\sqrt{v_3^2+\alpha }+v_3\right)\right)}{\sqrt{2} q_0 R_0
   \left(R_1^4-16\right)},~~~~~\alpha=\frac{N^2 q_0^2 R_0^6 \left(R_1^4-16\right){}^2}{2
   \left(R_1^4+8\right){}^2}.
\eeq
The behavior is similar for the exact solution,
although that is not SUSY on this cycle.
In all cases the cycle blows up in the UV and contracts to zero in the IR.
Here again, we should notice that the assumed range for the 
coordinate $v_3$ might imply that the cycle has a boundary.
We do not report about calibrated three-cycles or higher.

\section{Comments on the Quantum Field Theory.}\label{QFTaspects}   
In this section, we will study some aspects of   
the  four dimensional QFTs dual to the background    
we presented in eq.(\ref{metricxxx}). Comparisons with a suitable analysis   
for the solution after the non-abelian T-duality written in eqs.(\ref{iibmetricxxx})-(\ref{fluxesiibzz}), will be made when possible.   
   
We emphasize that the field theory dual to the Type IIA backgrounds
 is characteristically   
 non-local or `higher-dimensional'.   
This should not come as a surprise, as it was already observed in    
\cite{Itzhaki:1998dd},   
full decoupling of the gravity modes is not achieved    
for the case of flat D6 branes.   
We will make this point via the study of some observables   
that will be sensitive to the high energy properties   
of the QFT. We will analyse Wilson loops,    
with emphasis on its UV behavior. We will then study the entanglement   
entropy and central charge. Both observables will present signs of   
non-locality. We will also discuss the behaviour of Wilson, 't Hooft loops, 
domain walls and gauge couplings, when studied as IR effects.    
These observables are well-behaved for the solutions   
presented in this work. In other words, the dual QFT to   
our background in eq.(\ref{metricxxx}) or our new background 
in eqs.(\ref{iibmetricxxx})-(\ref{fluxesiibzz})---together    
with the solutions in Section \ref{explicitsolutions},   
behave as     
QFTs that at low energies show signs of the expected  four dimensional behaviour,
like  confinement and symmetry breaking, but    
need to be defined with a UV-cut off, or   
need a UV-completion.   
   
Various properties are 'inherited' (in a sense that will become
clear) by the   
new Type IIB solution that we have constructed. 
We will finally calculate the Page charges    
of this new solution. We will propose a possible quiver   
suggested by these charges.   

It will be clear by analising the backgrounds that the initial QFT,
corresponding to the compactified D6 branes has global symetries
given by $SU(2)\times SU(2)$, while the QFT dual to the Type IIB background
will only have $SU(2)$. This reduction of global symmetries 
(isometries, for the
dual backgrounds) is characteristic of non-Abelian T-duality.
\subsection{Some useful sub-manifolds}   
It will be useful for the analysis below, to define some sub-manifolds of the   
metric in eq.(\ref{metricxxx}).   
We can define then 
\bea   
\Sigma_3=[\theta,\varphi, \psi],\;\;\;\; \tilde{\Sigma}_3=[\tilde{\theta},\tilde{\varphi}, \psi],\;\;\;    
\hat{\Sigma}_3=[\theta=\tilde{\theta},\varphi=\tilde{\varphi}, \psi].   
\label{trescycles}   
\eea   
The volume element of each of these cycles is (we take $3A=\phi$),
\bea   
& & \sqrt{\det g_{\Sigma_3}}=16 \pi^2 (\alpha' g_s N)^{3/2} e^{\phi} h(b^2+a^2 g^2),\;\;\;\; \sqrt{\det g_{\tilde{\Sigma}_3}}=16 \pi^2  (\alpha' g_s N)^{3/2} e^{\phi} h a^2,\nonumber\\   
& & \sqrt{\det g_{\hat{\Sigma}_3}}=16 \pi^2  (\alpha' g_s N)^{3/2} e^{\phi} h\big(b^2+a^2 (g^2+1)\big).   
\label{threecycleszz}   
\eea   
We can see using the IR expansions that each of these cycles vanish at $r\to 0$   
and diverge as $r\to \infty$ for the explicit solutions presented in Section \ref{section: IIA}.   

If we consider the three-cycles after the 
non-Abelian T-duality, we have the submanifold
defined by the coordinates $(\tilde{\theta}, \tilde{\varphi}, v_2)$. 
This cycle is not calibrated and probably has a boundary in 
the coordinate $v_2$.
   
\subsection{Wilson and 't Hooft  loops.}   
The type IIA background in eq.(\ref{metricxxx}), ends in a  smooth way,   
with finite values for the combinations    
$   
F^2= g_{tt}g_{xx},\;\; G^2= g_{tt}g_{rr}.   
$   
This 
might suggest that the system confines    
as usual. But there are some subtleties.   
Indeed, when calculating the Wilson loop with the    
prescription of hanging a fundamental string from    
a  brane very far away in the UV of the geometry,   
we are assuming that this string will end on the D-brane satisfying    
the boundary condition of ending 'perpendicularly' to the brane.   
This is discussed, for example in   
\cite{Nunez:2009da}.    
Following the formalism in \cite{Nunez:2009da}, the boundary condition    
boils to defining   
$   
V_{eff}= \frac{F}{G}\sqrt{F^2-F_0^2}   
$   
and imposing that for large values of the radial coordinate   
$   
V_{eff}$ diverges.   
In our present case, $F^2=G^2= (\alpha' g_s N)^2 e^{\frac{4}{3}\phi}$ (we choose $\mu=1$). The value of    
$$   
V_{eff}\sim \sqrt{e^{4\phi/3}- e^{4\phi_0/3}}\alpha' g_s N   
$$   
is a finite constant for the semi-analytic solutions.   
This suggests,    
that the QFT needs to be UV-completed or be supplemented by a hard UV-cutoff   
which in turn suggests that the QFT   
is afflicted by the presence of an irrelevant operator.   
Conversely, one can consider the case in which the dilaton diverges at   
infinity, as   
described by eq.(\ref{exactedels}). In that case, the UV-boundary conditions are satisfied, but    
one will find that there is a minimal length-separation   
for the quark-antiquark pair. For $r_*$ close to the boundary    
$L_{QQ}(r_*)$ is finite, instead of vanishing.    
This indicates the presence of a minimal length   
in the dual QFT. Hence, some form of non-locality.   
In summary, regardless the solution we
choose, the high energy behaviour of the dual field theory
seems to be not the expected one for a 4-dimensional QFT.
   
Once assumed a UV-cutoff, the Wilson loop can be calculated.    
The QCD string tension is finite (suggesting confinement) and  given by,   
$$   
\sigma=\frac{1}{2\pi \alpha'} \sqrt{g_{tt}g_{xx}}|_{IR}=   
\frac{1}{2\pi\alpha'}e^{2A(0)}=   
\frac{(q_0 R_0)^2}{4\pi\alpha'} ~ .   
$$   
The components of the metric that enter this particular    
Wilson loop calculation are $g_{tt}, g_{xx}, g_{rr}$. These components   
are not changed by the non-Abelian T-duality.    
We should then expect that the comments above should be valid also for the   
QFT dual to the background in eq.(\ref{iibmetricxxx}).

In contact with the discussion on the dynamical 
character of the $SU(2)$-structure, notice that this is a consequence of
the deformation of the space associated with the confining behavior. 
Relations of this kind have been reported in \cite{Gaillard:2013vsa}.
\subsubsection{'t Hooft loops.}   
In a very similar way as described above, we could wrap a D4 brane    
on any of the three-cycles   
in eq.(\ref{trescycles}) and extend the brane on $[t,x_1]$, to form    
a magnetic string-like object. We propose that this object
computes the 't Hooft loop in the QFT.   
On the type IIA side, let us consider    
the different three-manifolds  in    
eq.(\ref{trescycles}),   
we will have that the effective tension of the 't Hooft string-like object is   
\bea
& & \frac{T_{eff,\Sigma_3}}{16\pi^2 T_{D4}(\alpha' g_s N)^{3/2}}= e^{5A-\phi}h (b^2+ a^2 g^2)|_{r=0}.   \nonumber\\
& & \frac{T_{eff,\tilde{\Sigma}_3}}{16\pi^2 T_{D4}(\alpha' g_s N)^{3/2}}= e^{5A-\phi}h a^2|_{r=0}.   \nonumber\\
& & \frac{T_{eff,\hat{\Sigma}_3}}{16\pi^2 T_{D4}(\alpha' g_s N)^{3/2}}= e^{5A-\phi}h \big(b^2+ a^2 (g^2+1) \big)|_{r=0}.   \nonumber
\eea    
Notice that all these present a vanishing tension--hence screening--   
of the monopole-antimonopole pair.   
Again, the behavior of this low energy 
observable is in line with the expected.   
   
We can define a screened magnetic string in the Type IIB picture.
To do so, we will use the two cycle described below
eq.(\ref{eq:2cy1})
and wrap a D3 brane on it, also extending the brane on
the  two directions  $(t,x_1)$.
For the effective tension we will get,
\beq
\frac{T_{eff}}{T_{D3}}= e^{A-\Phi}\int d\theta d\varphi \sqrt{\det[g+B]_{\Sigma_2}}|_{r=0}.
\eeq
We observe using the asymptotics associated with this cycle
a tensionless magnetic string or conversely, 
a 'screened' force between a pair of monopoles, as expected.   
Let us move to study another IR-observable.   
\subsection{Domain Walls}   
In our Type IIA  geometry of eq.(\ref{metricxxx}), there is a natural two-cycle   
defined by   
$$   
\Sigma_2=[\theta=\tilde{\theta}, \varphi=\tilde{\varphi}],   
$$   
for some fixed value of the angle $\psi=\psi_0$, which is SUSY in the IR.    
   
The objects of potential interest to represent Domain Walls,    
are D4 branes that wrap the two-cycle   
above and that extend on the Minkowski directions $(t, x_1,x_2)$.   
If this object has finite tension, then it may act    
as a Domain Wall, separating different vacua.   
Let us study the object in more detail.   
   
The induced metric (for constant radial coordinate and constant    
angle $\psi$) is,   
\beq   
ds_{ind,st}^2= e^{2A}\Big[\mu dx_{1,2}^2    
+ \alpha'g_s N \Big( b^2+ a^2(g^2+1 +2 g \cos\psi)  \Big)   
(d\theta^2+\sin^2\theta d\varphi^2)      \Big].   
\eeq   
So, the action of the object (choosing $\mu=1$)   
is,   
\beq   
S=-T_{eff}\int d^3 x, \;\;\;\;    
T_{eff}=
16\pi^2 e^{5A-\phi}(\alpha'g_s N)T_{D4}\Big( b^2+ a^2(g^2+1 +2 g \cos\psi_0)  \Big)|_{r=0}.   
\eeq   
We can use the IR expansions of eq.(\ref{eq:Dat0-bis}),    
to check that this object has a    
constant tension in the far IR of the geometry. 
If we follow the logic presented in   
\cite{Acharya:2001dz} and    
add a    
gauge field ($a_1$, with curvature $f_2=da_1$) on   
 the Minkowski part of   
the world volume of the brane    
This will create a Wess-Zumino term of the form   
\beq   
S_{WZ}=T_{D4}\int C_1 \wedge f_2\wedge f_2=-T_{D4}\int d\theta d\varphi F_2    
\int d^3 x f_2\wedge a_1.   
\eeq   
Using that on the particular    
cycle $F_2=-2 N \sin\theta d\theta \wedge d\varphi$,   
we have induced a Chern-Simons term.       
These domain walls, should separate vacua coming from the breaking of   
some global (discrete) symmetry, see  
\cite{Gursoy:2003hf}.   
   
After the non-Abelian T-duality, we can define Domain Walls
by using the calibrated one-cycle defined around eq.(\ref{1-cycleiib})
and extend a D3 brane on the $(t, x_1, x_2)$ directions, also wrapping the one-cycle
parametrised by $v_3$.
We will have a simple induced metric
\beq
ds^2_{D3}= e^{2A}(-dt^2 + dx_1^2+ dx_2^2) + d\Sigma_1^2.
\eeq
The Action and  effective tension of this object will be given by,
\bea
& & S_{D3}= -T_{D3}e^{3A-\Phi}\sqrt{\det g_{\Sigma_1}} \int dv_3 \int d^{2+1}x,\nonumber\\
& & T_{eff}=T_{D3}\int dv_3 
e^{3A-\Phi}\sqrt{\det g_{\Sigma_1}}|_{r=0}.
\eea
Notice that imposing that the Domain Wall has a finite tension implies a finite range of values (or periodicity) for the coordinate $v_3$.
Here again, like when we restricted the range 
of $v_2$ to avoid singularities---
see around eq.(\ref{nonsingularitycondition}),
 we find that a  'physical' requirement implies 
conditions on the range of coordinates.
These conditions are not imposed by non-Abelian 
T-duality when thought as a solution 
generating technique in supergravity.
In other words, the periodicty of the corodinate $v_3$ is being {\it imposed}
by the requirement that the domain-wall objects in the dual QFT have finite
tension. This type of requirements may give hints about the 
Type IIB geometry we have generated.

We can also turn on a gauge field ${\cal A}$ 
with curvature ${\cal F}_2 $ on the  $R^{1,2}$ directions. The Wess-Zumino term will read
\beq
S_{WZ}= \Big(T_{D3}\int_{v_3} F_1|_{r=0} \Big)\int d^{2+1}x {\cal A}_1 \wedge {\cal F}_2 =\kappa \int d^{2+1}x {\cal A}_1 \wedge {\cal F}_2.
\eeq
Using that the Ramond form $C_0= 2N(K+1) v_3$---see below eq.(\ref{xxyy})--- implies that 
the 'charge' of the Domain Wall (or the coefficient of the Chern-Simons term induced on it) is
\beq
\kappa= 2N(K(0)+1)\oint dv_3. \;\;\; K(0)=1
\eeq

Let us move now to the definition of a gauge coupling.
\subsection{A gauge coupling}   
We can define the gauge coupling of the QFT, by   
wrapping a D6 brane on any of the three-cycles in eq.(\ref{trescycles}).    
We turn on a gauge field on the brane   
(for the argument, it is enough to turn on just $F_{tx_1}$),   
and we also turn on   
a pure gauge $C_3$-field of the form   
$$   
C_3=\frac{k}{16\pi^2}\sin\tilde{\theta}d\tilde{\theta}\wedge d\tilde{\varphi}   
\wedge d\psi,   
$$   
we will have, for the cycle   
$\tilde{\Sigma}_3$ in eq.(\ref{trescycles}) \footnote{We found that this cycle fails to be calibrated, in far UV, by a factor of $1/2$.}    
that the induced metric and Born-Infeld-Wess-Zumino-action are (we use $3A=\phi$),   
\bea   
& & ds_{\tilde{\Sigma}_3}^2= e^{2\phi/3}\Big[\mu dx_{1,3}^2    
+\alpha'g_s N\Big(  a^2(\omega_1^2 +\omega_2^2)+   
h^2 \omega_3^2 \Big) \Big], \nonumber\\   
& & S_{BIWZ}= - T_{D6}\int e^{-\phi}\sqrt{-\det[g_{ab}+2\pi \alpha'    
F_{ab}]} + T_{D6}   
\int C_7 + C_3\wedge F_2\wedge F_2.   
\\   
& & S_{BIWZ}\sim    
- T_{D6}(\alpha' g_s N_c)^{3/2} 16 \pi^2 \int e^{\frac{4\phi}{3}}\mu^2    
h a^2(    
1- \frac{1}{2 \mu^2} e^{-4\phi/3}4\pi^2 \alpha'^2 F_{\mu\nu}F^{\mu\nu} )   
\nonumber\\   
& & +    
T_{D6}k \int F_2\wedge F_2 + T_{D6}\int C_7.\nonumber   
\label{gaugecouplingBI}   
\eea   
where the last contraction $F_{\mu\nu}F^{\mu\nu}$    
is in Minkowski space and    
we have expanded for small   
field strengths (equivalently for small values of $\alpha'$).   
This leaves us with a  gauge coupling of the form,   
\beq   
\frac{1}{g_{YM}^2 N_c}= (g_s N)^{1/2}\frac{a^2 h}{2 \pi^4}.   
\label{gaugecouplingsigma3}   
\eeq 
with asymptotic behaviour as $r\rightarrow \infty$,
\begin{align}
	\frac{1}{g_{YM}^2 N_c}&\sim \frac{(g_s N)^{1/2}}{2 \pi^4}\left( \frac{r^3}{18} - \frac{1}{2} q_1 R_1\  r^2 + \frac{3}{16} (3 R_1^2 + 8 q_1^2 R_1^2)\  r - 
\frac{3}{16} (9 q_1 R_1^3 + 8 q_1^3 R_1^3) +\frac{ 801 R_1^4}{256 }\frac{1}{r}+\cdots\right)
\end{align}
and as	$r\rightarrow0$
\begin{align}
	\frac{1}{g_{YM}^2 N_c}&\sim\frac{ (g_s N)^{1/2}}{2\pi^4}\left(\frac{r^3}{8} + \frac{(-8 - q_0^2)}{768 R_0^2}\ r^5 + \frac{(1792 - 208 q_0^2 - 93 q_0^4)}{(737280 R_0^4)}\ r^7+ \cdots\right)\\
\end{align}
Notice that there is no effect of the rescaling by $\mu$.   
This is expected, because this defines a a four-dimensional gauge coupling,   
that should be classically invariant under dilations.   
   
We can run this calculation for the other three-cycles    
defined in eq.(\ref{trescycles})   
and get analogous expressions.   
All these expressions present a divergent gauge coupling in    
the IR---in the solution   
of eq.(\ref{exactedels}) it diverges at $\r=a$--- while    
vanishing in the far UV.   
This should not be taken as a sign that the QFT is weakly    
coupled in the far UV. Indeed, these QFTs contain also    
superpotential couplings that make the whole system strongly interacting.   
This is in agreement with the dual spacetimes 
being weakly curved and trustable   
in the far UV.   

After the non-Abelian T-duality, we can
define a gauge coupling in the type IIB dual by using D5 branes; extend them on $R^{1,3}$ and wrapping
the calibrated two cycle defined below eq.(\ref{eq:2cy1}).
We should also turn on a gauge field on the $R^{1,3}$ directions 
and also consider the projection of the NS $B_2$ field
on the two-cycle. We find that this gauge coupling reads,
\beq
\frac{1}{g^2}= 4\pi^2 \alpha'^2 T_{D5} e^{-\Phi}\int \sqrt{\det[g_{\Sigma_2}+B_2]}
\eeq
Using the asymptotics associated with the cycle above, 
we see that this gauge coupling 'confines' in the IR and 
vanishes in the far UV. 
The Wess-Zumino term for this D5 brane should define the $\Theta$-angle.
   
In summary, we see that these observables, behave in the far IR   
as expected for a confining four dimensional QFT. Nevertheless, the Wilson loop   
indicates the need for a UV-completion. Below, we will briefly discuss   
another observable showing the same need for UV-completion.   
\subsection{Central Charge and Entanglement Entropy}   
A couple of quantities that characterise nicely the QFT dual to a geometry   
are the central charge and entanglement entropy of the QFT.   
These quantities have been studied in many different papers. Let us quote   
a couple of original references   
\cite{Girardello:1998pd}, \cite{Ryu:2006bv}.   
   
We will follow    
the systematic treatment summarised   
in    
\cite{Klebanov:2007ws}. Consider a metric of the form,   
\beq   
ds_{st}^2= \alpha \beta dr^2 + \alpha dx_{1,d}^2 + g_{ij}dy^i dy^j,   
\eeq    
we can compute the following quantities in our generic background of    
eq.(\ref{metricxxx})   
\bea   
& & V_{int}= \int d^{8-d}y \sqrt{\det[g_{ij}]}=    
(4\pi)^3 b^2 a^2 h (\alpha' g_s N)^{5/2} e^{5\phi/3},\nonumber\\   
& & \alpha = \mu e^{2A},\;\;\;\; \beta= \frac{\alpha' g_s N}{\mu},   
\;\;\; d=3,\\   
& & H= e^{-4\phi}V_{int}^2\alpha^d= (4\pi)^6 \mu^3 b^4 a^4 h^2 (\alpha' g_s    
N)^{5}   
e^{16 A-4\phi},\nonumber\\   
& & ds_5^2= \kappa [dx_{1,3}^2 + dr^2],\;\;\;\; \kappa^3=H   
\nonumber   
\eea   
This implies that the central charge is given by,   
\beq   
c\sim 27 N^{3/2} \frac{H^{7/2}}{(H')^3}.   
\eeq   
The UV and IR behavior of the central charge  for the solution with stabilized dilaton is
\begin{align}
	r&\rightarrow \infty\nonumber\\
	&\log(c)\sim -8 \log(1/r) + \frac{7}{2} \log(\frac{R_1^2}{11664}) - 3 \log(\frac{5 R_1^2}{5832})-\log(2) - 
	24 \frac{q_1 R_1}{r} +\frac{ (\frac{297 R_1^2}{40} - 36 q_1^2 R_1^2)}{r^2}+\cdots\nonumber\\
	r&\rightarrow 0\nonumber\\
	&\log(c) \sim 6 \log(r) + \frac{7}{2} \log\frac{ q_0^2 R_0^6}{64} - 
	3 \log\frac{3 q_0^2 R_0^6}{32}-\log(2) + \frac{-4 + q_0^2}{24 R_0^2} r^2+\cdots
\end{align}

\begin{figure}\label{fig:central-charge}	   
	\begin{center}   
\includegraphics[height=2in]{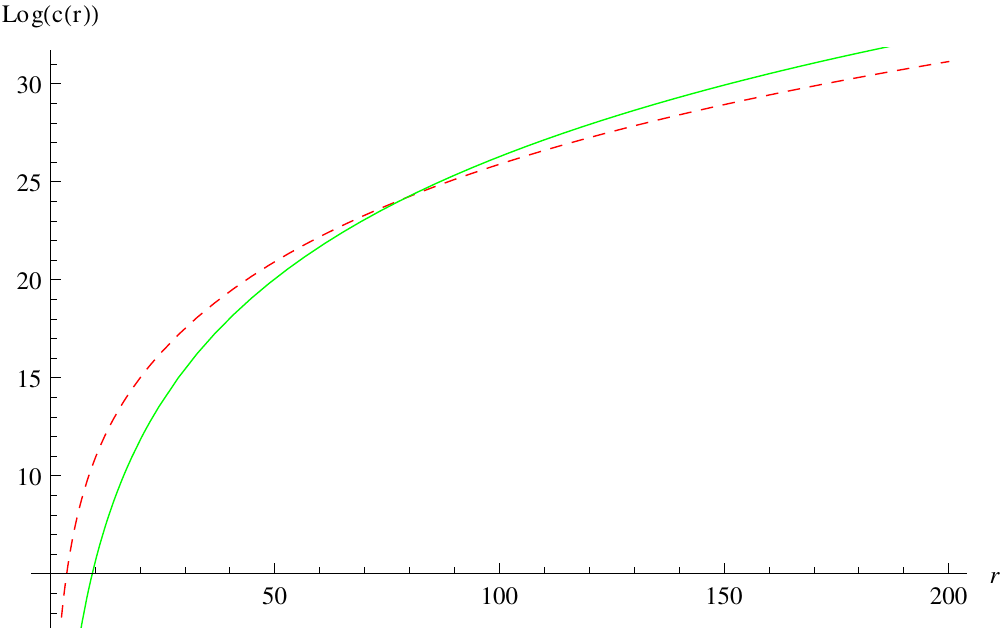}   
\caption{The central charge for a numerical solution with stabilized dilaton (red dashed curve) and for the exact solution with linear dilaton (green curve). }   
\end{center}   
\end{figure}

For comparison, we note that  the central charge of the exact solution is, in the UV, $\log (c_{exact})\sim \log(\frac{r^9}{2239488 \sqrt{3}})$. In Figure (\ref{fig:central-charge})    
we plot the central charge for a numerical solution with stabilized    
dilaton and for the exact solution with linear dilaton.    
   
If we      
calculate the central charge after the non-abelian T-duality    
using the background   
of eq.(\ref{iibmetricxxx}), we follow \cite{Klebanov:2007ws}   
and write the relevant quantities are,   
\beq   
\alpha=e^{2A}\mu, \;\;\; \beta= \frac{\alpha' g_s N_c}{\mu},\;\;\;   
V= \int d\theta d\varphi d\psi dv_1 dv_2 e^{-2\hat{\Phi}}\sqrt{g_{int}}.   
\eeq   
and the we will have   
$$   
H=V^2 \alpha^3, \;\;\; c\sim \frac{H^{7/2}}{(H')^3}.   
$$   
Following the algebra, one gets   
$$   
c_{new}=\pi{\cal N} c_{old}.   
$$   
Where ${\cal N}$ is an radius (energy) independent factor.   
Then, the central charges of the original and    
T-dual solutions differ by a constant with no much dynamical content.    
This can be traced to the    
invariance under NATD of the quantity $\sqrt{g_{initial}} e^{-\phi_{initial}}$,    
being equal, up to a    
Fadeev-Popov like factor to the same quantity in the    
dual background.   
This is explained in \cite{Itsios:2013wd}. The    
Fadeev-Popov factor is associated    
with the scale independent number ${\cal N}$ above. This central charge   
and the    
entanglement entropy    
described below are two observables whose behavior is 'inherited'   
by the non-Abelian T-dualised background QFT pair.   
\subsubsection{Entanglement Entropy.}   
We  now turn to the entanglement entropy. Consider a boundary region     
$\mathbb{R}^{d-1} \times \mathcal{I}_L$ where $\mathcal{I}_L$ is a line segment of length L. We calculate the entanglement entropy  following   
\cite{Klebanov:2007ws} and   obtain,    
\bea   
& & L(r_*)=2 \sqrt{H(r_*) N}\int_{r_*}^{\infty}    
\frac{dr}{\sqrt{H(r)- H(r_*)}}, \label{eq:Lrs}\\   
& & S_{conn}-S_{disc}\sim \int_{r_*}^{\infty}dr \sqrt{H}   
\Big[\frac{\sqrt{H}}{\sqrt{H-H(r_*)}} -1 \Big]- \int_{r_0}^{r_*} dr\sqrt{H}.   
\eea    
Evaluating (\ref{eq:Lrs}) using the numerical solutions    
with stabilized dilaton found in  Section \ref{explicitsolutions}    
we can show that $L(r_*)$  grows indefinetely and   has not    
a maximum value. The non-existence of a maximum and hence the   
absence of double-valuedness for $L(\r_*)$, suggests the absence of a    
first order phase transition in the entanglement entropy.   
This falls within the description of \cite{Barbon:2008ut}   
for the entanglement entropy of non-local QFTs. Same behavior will present   
the background of eq.(\ref{iibmetricxxx}).   
   
A tricky point that should not confuse the diligent reader   
is that if a UV cutoff is imposed on the geometry, numerically   
a double valuedness of $L(r_*)$ is obtained 
and correspondingly, a first order transition   
in the entanglement entropy will be observed. But a 
more detailed analysis   
will show that changing the position of the cutoff,    
moves also the position of the maximum   
of the separation $L(r_*)$ and the maximum of the    
phase transition. Hence, this is a cutoff effect and should 
perhaps be taken as   
non-physical. 
The resolution
is that a  cutoff in the radial direction is needed to solve
some stability problems in the configurations that compute
the Entanglement Entropy. At the same time a Volume-law for the
divergent part of the Entanglement Entropy will take place.
A more detailed analysis of these issues 
appears in \cite{Kol:2014nqa}. 
 
\subsection{Page Charges}   
Finally, we will study some global quantities in 
the QFT that are defined using the
background of eqs.(\ref{iibmetricxxx})-(\ref{xxyy}). 
Following \cite{Marolf:2000cb} we write   
some given currents at constant radial position,   
\beq\label{eq pagecurrents}   
\begin{array}{l l}   
\vspace{3 mm}   
\star\mathcal{J}^{Page}_{D7}&=d F_1,\\   
\vspace{3 mm}   
\star\mathcal{J}_{D5}^{Page}&=d(F_3-B_2\wedge F_1)\\   
\vspace{3 mm}   
\star\mathcal{J}_{D3}^{Page}&=d(F_5-B_2\wedge F_3    
+\frac{1}{2} B_2\wedge B_2 \wedge F_1).   
\end{array}   
\eeq   
In terms of these we can define three Page charges,   
\beq   
Q^{page}_{D7}=\frac{1}{2\kappa_{10}^2T_{D7}}\int_{V_2}\star\mathcal{J}^{Page}_{D7},   
~~~Q^{page}_{D5}=\frac{1}{2\kappa_{10}^2T_{D5}}\int_{V_4}\star\mathcal{J}^{Page}_{D5},   
~~~Q^{page}_{D3}=\frac{1}{2\kappa_{10}^2T_{D3}}\int_{V_6}\star\mathcal{J}^{Page}_{D3}.   
\eeq   
where $V_{9-p}$ is the transverse space of the corresponding Dp brane.    
Using Stokes theorem these may be expressed as integrals over three    
compact spaces. Notice that we    
demand that $v_2$ and $v_3$ 
are compact to have these charges well-defined.    
Let us propose the following cycles at constant radius (the coordinates not
mentioned are kept at constant values),   
\beq   
\Sigma_1= (v_3), ~~\Sigma_3=(\tilde{\theta},\tilde{\varphi},v_2=v_3),~~   
~~\Sigma_5=(\tilde{\theta},\tilde{\varphi},v_2,v_3,\psi)   
\label{cicloszz}   
\eeq   
Then the Page charges may be expressed as   
in the paper \cite{Benini:2007gx}  by the following quantities,   
\bea   
& & Q_{D7}=\frac{1}{4}\int F_1, \;\;\;\;    
Q_{D5}=\frac{1}{16\pi^2}\int F_3- B_2\wedge F_1\nonumber\\   
& & Q_{D3}=\frac{1}{64\pi^4}\int F_5- B_2\wedge F_3 +\frac{1}{2}   
B_2\wedge B_2\wedge F_1 .\nonumber\\   
\eea   
We then get that the relevant quantities are,
\beq
\begin{array}{ll}
&F_1= -2N (K+1)dv_3\\[3 mm]
&F_3-B_2\wedge F_1= \sqrt{2}N(K-1)
\sin\tilde{\theta}\big(v_2dv_2+v_3dv_3\big)
\wedge d\tilde{\theta}\wedge d\tilde{\varphi}\\[3 mm]
&F_5-B_2\wedge F_3 +\frac{1}{2}B_2\wedge B_2 \wedge F_1= 0.
\end{array}
\eeq
Performing explicitly the integrals, we get
\bea   
Q_{D7} =- N \hat{A} (K+1),\;\;   
Q_{D5} =N (K-1) \hat{B},\;\;Q_{D3} =0.
\eea   
Importantly, we have imposed that the range of the coordinates
$v_2,v_3$ is finite. We have defined them as periodic with 
periodicity of the coordinate $v_3$ being $\hat{A}$   
and that for $v_2$ being $\hat{B}$, according to,   
\bea   
\hat{A}= \frac{1}{2}\int dv_3,\;\;\; \hat{B}=\frac{1}{\sqrt{2}\pi}
\int v_2 dv_2   
\eea   
The integrals are performed over the range of those variables $v_2, v_3$.   
If the manifolds in eq.(\ref{cicloszz}) were strictly 'cycles'
the Page charges above defined should all be quantised, see our comments
above about the presnces of boundaries in these submanifolds.
We will then impose
a quantisation on a combination of $Q_{D7}$ and $Q_{D5}$.
Indeed, we can form the combination,  
\bea   
Q_{int}=-(Q_{D7}+ Q_{D5})
\eea   
If we impose that the periods $\hat{A}, \hat{B}$
are equal and integer, we have defined a quantised quantity $Q_{int}$. 
This together with $Q_{D3}$, suggest a  situation reminiscent 
of the Klebanov-Strassler QFT, with two gauge groups and one of the
Page charges (that associated with D3 branes), vanishing.

This suggests that we are dealing with a two-nodes quiver, 
plus some bifundamental matter. It is certainly not the KS-field theory.
We leave for future studies to describe the precise matter 
content and interactions of the bifundamental matter.

\section{Conclusion and Future Directions.}\label{conclusionsxx}  
Let us start by briefly summarising what we have done in this paper.
We started with backgrounds in M-theory, reduced them to Type IIA, wrote the 
conditions for these backgrounds to preserve minimal SUSY in four dimensions
(this was material already present in the bibliography).
The first piece of new material consisted in explicitly
solving the differential equations with a careful 
numerical integration that used as boundary conditions 
the asymptotic solutions,
obtained analytically by solving (asymptotically) the BPS system. This is why 
we called our solutions 'semi-analytical'.
We then studied the transition between $G_2$ structure (in eleven dimensions)
to $SU(3)$ structure in Type IIA. We constructed explicit expressions for 
the potential and calibration forms.

Then, we performed a non-Abelian T-duality on this Type IIA background.
We obtained a family of backgrounds in Type IIB with all Ramond and 
Neveu-Schwarz forms turned on. This is a {\it new} family of solutions.
We established its
$SU(2)$-dynamical structure, pure spinors, calibration forms
and found some calibrated cycles.
Restrictions on the range of the T-dual
coordinates were imposed, by requiring the 
smoothness of the generated space and the good behavior of field theoretical observables.

After that, we moved into the study of the 
correspondence between the family of Type IIA 
solutions and its dual QFT, also
extending the study of various observables to the QFT's dual to the new family 
of IIB backgrounds. In this line, we made clear that the QFTs are non-local
and in the need of a UV-completion (this is specially clear from the 
behaviour of the Wilson loop and central charges 
at high energies). On the other hand, 
observables relevant to the IR dynamics show the expected four-dimensional 
behaviour. Finally, based on global charges, we loosely proposed
a possible two-nodes 
quiver describing the QFT dual to the new Type IIB background.
Notice that in the logic we are advocating, the background is {\it
defining the QFT}
via its observables at strong coupling.

A couple of points emerged as specially interesting from the previous study.
If we impose that some physical observables of the 
QFT dual to our new background
behave as expected, this in turn imposes 
constraints on the new coordinates 'after the duality'. We also restricted
the range of one of the dual coordinates $v_2$ in order to avoid singularities. This is not free of 
ambiguities,
unlike the restriction imposed on $v_3$ to be periodic, such that the domain wall 
charge is quantised. 

 .
These new coordinates originally play the 
role of Lagrange multipliers in the sigma model Action.
Working at the genus-zero level in the sigma model 
gives no information on the periodicity (or not),
of such new coordinates. It is quite nice to find 
some conditions imposing the good-behaviour
of the dual QFT. 

It is also quite interesting to have found 
an $SU(2)$-dynamical structure in Type 
IIB for a solution preserving four supercharges.
It is our understanding that such backgrounds are  
not  easy to come by. The technique
presented here suggests a way of generating 
these and other backgrounds with similar features.

What could be nice to study (and at the same time feasible)?
It seems natural to explore further 
the quiver structure of the QFT.
It would be interesting to search in our backgrounds
other well defined strong coupling effects and observables 
believed to appear in those QFTs. In this way, try to make sharp the
dual QFT (matter content, superpotential, etc).

Even more interesting, but perhaps more difficult, 
would be to find a UV completion to our 
Type IIB dual QFT. Thinking about the lines of the papers 
\cite{Maldacena:2009mw}  one may find a way to
transform our system into one presenting $AdS_5$-like asymptotics. 

Extending our results
to examples in $2+1$ and $1+1$ dimensions seems a natural way to proceed.
All these studies mentioned above will give important 
clues into the understanding
of non-Abelian T-duality.

\section{Acknowledgments:} We wish to thank various physicists    
for nice discussions: David Andriot, Jerome Gaillard, Yolanda Lozano,
 Michela Petrini, Daniel Thompson,   Diego Rodriguez-Gomez, Leo Pando-Zayas, 
Daniel Schofield, Kostas Sfetsos, Alessandro Tomasiello, Michael Warschawski, Alberto Zaffaroni.  
E.C. acknowledges support of CONACyT grant CB-2008-01-104649 
and of the  National Science Foundation under Grant PHY-1316033.
Niall Macpherson is supported by an STFC studentship. 
He is greatful for
the warm hospitality extended by hep-th group 
at Oviedo University, where
part of this work was performed 
and to the COST Action MP1210 "The string
theory Universe", for funding a research visit there.    
This paper was started while Carlos Nunez was a Feinberg Foundation
Visiting Faculty Program Fellow, he thanks the hospitality 
extended at Weizmann Institute
and The Academic Study Group for the Isaiah Berlin Travel award.

\appendix   
\section{Appendix: On numerics}\label{appendix-numerics}  
Our goal is to numerically find some particular solutions of the  equations 
\bea\label{eq:Deqnsx}   
\dot{a} = -\frac{c}{2a} + \frac{a^5 f^2}{8 b^4 c^3}, & \;\; &   
\dot{b} = -\frac{c}{2b} - \frac{a^2 (a^2-3c^2)f^2}{8b^3c^3}, \nn \\   
\dot{c} = -1+\frac{c^2}{2 a^2}+\frac{c^2}{2 b^2}-\frac{3 a^2   
f^2}{8b^4}, & \;\; &   
\dot{f} = -\frac{a^4 f^3}{4 b^4 c^3}.  
\eea   
In general this system will have four integration constants. We can find series solutions of these equations as $r\rightarrow 0$ and  choose the  the zeroth order term in each expansion to be the independent parameter. Thus, generally  the IR expansions will have the form,
\be
a(r)\sim a_0 + a_1(a_0,b_0,c_0,f_0) r + a_2(a_0,b_0,c_0,f_0) r^2 + a_3(a_0,b_0,c_0,f_0) r^3 + \cdots
\ee
and similar expressions for all the other functions. 
However, we are interested in solutions dual to a 4 dimensional field theory, thus we want the 3-cycle that the D6 brane wraps to shrink to zero  as $r\rightarrow0$. From  the IIA metric \eqref{metricxxx} we see that  this requirement fixes $a_0=0, \ c_0=0$ and we are left with only two independent parameters in the IR, $b_0$ and $f_0$ that we label $R_0$ and $q_0 R_0$ respectively. Similarly, in the UV generically we have 4 independent parameters but since we want solutions with a stabilized dilaton, we set the coefficient of the linear term in the dilaton expansion to zero and are left with three 
independent parameters ${R_1,q_1,h_1}$ in terms of which a UV solution to arbitrary order can be found. 

To find numerical solutions we have the choice of starting in the  IR and integrate forward or start in the UV and integrate backwards. 
We choose to solve the equations of motion starting from the IR, using the IR expansions as boundary conditions. Our motivations for 
doing so are two-fold. First, the parameter space in the IR is smaller, $\{R_0, q_0\}$, than the one in the UV, $\{R_1,q_1, h_1 \}$, this facilitates the search of a solution with the required behavior. Second, the expansion of the equations of motion around $r=0$ is less 
computationally-intensive than the one around $r\rightarrow\infty$ allowing us to use very high order expansions as boundary conditions. More precisely, in our code we use IR expansions of the functions $a(r),\ b(r),\ c(r),\ f(r)$ up to order  ${\mathcal O}(r^{27})$ as boundary conditions. 
By way of illustration,  we present here  the IR expansions  up to order  ${\mathcal O}(r^{13})$,
\begin{align}
a(r) &= \frac{r}{2} - \frac{{(2 + q_0^2)} r^3}{(
			288 R_0^2)} + \frac{(74 + 29 q_0^2 - 31 q_0^4) r^5}{(
			69120 R_0^4)} + \frac{(-7274 + 546 q_0^2 + 5043 q_0^4 - 2473 q_0^6) r^7}{(34836480 R_0^6)} +\nonumber\\
&\frac{(-2767396 + 2066644 q_0^2 + 1326639 q_0^4 - 2267840 q_0^6 + 761969 q_0^8) r^9}{(60197437440 R_0^8)} +  \frac{P_{10}(q_0) r^{11}}{(158921234841600 R_0^{10})} -\nonumber\\
	&\frac{ P_{12}(q_0) r^{13}}{(297500551623475200 R_0^{12})}\nonumber\\
      P_{10}(q_0)&=-1732820552 + 2661492292 q_0^2 - 
        714674162 q_0^4 - 1616450167 q_0^6 + 1494468524 q_0^8 - \nonumber\\
	& 388078387 q_0^{10}\nonumber\\
	P_{12}(q_0)&=809180302184 - 1936619471316 q_0^2 + 
      1686929485098 q_0^4 + 13678188077 q_0^6\nonumber\\
      & - 1046636256642 q_0^8 + 
      694139577405 q_0^{10} - 148147907158 q_0^{12}
      \end{align}

      \begin{align}
	      b(r)&= R_0 - \frac{(-2 + q_0^2) r^2}{(
		      R_0 16)} - \frac{(13 - 21 q_0^2 + 11 q_0^4) r^4}{(1152 R_0^3)}  + \frac{(3268 - 
		      8866 q_0^2 + 9149 q_0^4 - 3209 q_0^6) r^6}{(1658880 R_0^5)} \nonumber\\
	      &  +  \frac{Pb_{8}(q_0)r^8}{(557383680 R_0^7)}   + \frac{Pb_{10}(q_0)r^{10}}{(1203948748800 R_0^9)}+ 
	      \frac{Pb_{12}(q_0)r^{12}}{(3814109636198400 R_0^{11})}\nonumber\\
      Pb_{12}(q_0)&=-96075595496 + 555977381336 q_0^2 - 
     1393711678048 q_0^4 + 1890154422552 q_0^6\nonumber\\
     &- 1451154850145 q_0^8 + 
     596013842074 q_0^{10} - 102144488257 q_0^{12} \nonumber\\
      Pb_{10}(q_0)&=120346756 - 576435426 q_0^2 + 
     1165086146 q_0^4 - 1196194108 q_0^6 + 617593365 q_0^8 - 
     127804976 q_0^{10} \nonumber\\
     Pb_{8}(q_0)&=-235082 + 885868 q_0^2 - 1355526 q_0^4 + 
     938210 q_0^6 - 244621 q_0^8
      \end{align}

      \begin{align}
	      c(r) &= -\frac{r}{2} + \frac{(8 - 5 q_0^2) r^3}{(
		      288 R_0^2)} - \frac{(232 - 353 q_0^2 + 
		      157 q_0^4) r^5}{(34560 R_0^4)} +\frac{(31168 - 76440 q_0^2 + 68637 q_0^4 - 21286 q_0^6) r^7}{(
		      17418240 R_0^6)}\nonumber\\
	      &+\frac{ Pc_{10}(q_0) r^{11}}{(39730308710400 R_0^{10})} + \frac{Pc_{8}(q_0) r^9}{(15049359360 R_0^8)} + 
	      \frac{ Pc_{12}(q_0)r^{13}}{(74375137905868800 R_0^{12})}\nonumber\\
     Pc_{10}(q_0)&=5716032512 - 24717750400 q_0^2 + 
       44863517744 q_0^4 - 41761366916 q_0^6 + 19753037956 q_0^8\nonumber\\  
       & \quad\quad - 3779455283 q_0^{10}\nonumber\\
       Pc_{8}(q_0)&=-7527424 + 
     25507072 q_0^2 - 34570320 q_0^4 + 21451291 q_0^6 - 
     5080615 q_0^8\nonumber\\
      Pc_{12}(q_0)&= -3137711476736 + 16500424668672 q_0^2 - 
     37556084710560 q_0^4 + 46609546892530 q_0^6 -\nonumber\\ 
     & \quad \quad33023463748437 q_0^8 + 12612429685326 q_0^{10} - 
     2023272290207 q_0^{12}
     \end{align}

     \begin{align}
	     f(r)&=q_0 R_0 + \frac{q_0^3 r^2}{
		     R_0 16} + \frac{q_0^3 (-14 + 11 q_0^2) r^4}{(1152 R_0^3)} +\frac{ 
		     q_0^3 (2152 - 3473 q_0^2 + 1492 q_0^4) r^6}{(829440 R_0^5)}\nonumber\\
	     &+ 
	     \frac{Pf_{8}(q_0)r^{10}}{(1203948748800 R_0^9)}
  +
  \frac{Pf_{6}(q_0) r^8}{(
	  557383680 R_0^7)} +\frac{ Pf_{10}(q_0)r^{12}}{(238381852262400 R_0^{11})}\nonumber\\
      Pf_{8}(q_0)&= q_0^3 (170283008 - 568700672 q_0^2 + 744979116 q_0^4 - 
    446064434 q_0^6 + 102094739 q_0^8)\nonumber\\ 
 Pf_{6}(q_0)&= q_0^3 (-329536 + 813288 q_0^2 - 705252 q_0^4 + 210349 q_0^6)\nonumber\\
  Pf_{10}(q_0)&=q_0^3 (-8376443008 + 35383047296 q_0^2 - 
      62151718900 q_0^4 + 55981055275 q_0^6 - 25662630839 q_0^8 +\nonumber\\ 
     & \quad\quad 4767879802 q_0^{10})    
   \end{align}

Using 40-digit \textbf{WorkingPrecision} in \textbf{ NDSolve}, Mathematica 8, we generate, using the IR expansions as boundary conditions,  solutions that extend in the UV. 
   We observe that not for all values of $\{R_0, q_0\}$ we get solutions with stabilized dilaton. Thus, the behavior of the dilaton serves as a first indication of a potential solution with the required UV behavior. We use  UV expansions up to order   ${\mathcal O}(1/r^{9})$ for all the functions. We show here, as an example, the UV expansion for $a(r)$. 
\begin{align} \label{eq:sol-UV-app}   
	a(r)&=	\frac{r}{\sqrt{6}}-\frac{\sqrt{3}  {q_1} R_1}{\sqrt{2}}+\frac{21 \sqrt{3}   
		{R_1}^2}{\sqrt{2}\  16 r}+\frac{63 \sqrt{3}   {q_1}   {R_1}^3}{\sqrt{2}\ 16 r^2}+\frac{9   
		\sqrt{3} \left(672   {q_1}^2+221\right)   {R_1}^4}{\sqrt{2}\ 512 r^3}+\nonumber\\   
  & \frac{81 \sqrt{3}   {q_1} \left(224   {q_1}^2+221\right)   {R_1}^5}{\sqrt{2}\  512   
   r^4}+\frac{\sqrt{3} \left(2048    {h_1}+1377 \left(768   {q_1}^4+1632   
		   {q_1}^2+137\right)   {R_1}^6\right)}{\sqrt{2}\  8192 r^5}+ \nn\\
		 & +\frac{3 \sqrt{\frac{3}{2}}
			 q_1 R_1 (10240 h_1 + 81 (11645 + 68000 q_1^2 + 22272 q_1^4) R_1^6)}{
			 8192 r^6}\nn\\   
		 &+	\frac{ 27 \sqrt{\frac{3}{2}} R1^2 (8192 h_1 (27 + 560 q1^2) + 
   27 (583399 + 17765952 q_1^2 + 68376576 q_1^4 + 
   20299776 q_1^6) R_1^6)}{3670016 r^7}\nn\\
   &+ \frac{27 \sqrt{\frac{3}{2}} q_1 R_1^3 (8192 h_1 (81 + 560 q_1^2) + 
   81 (583399 + 7332032 q_1^2 + 20121600 q_1^4 + 
   5849088 q_1^6) R_1^6)}{524288 r^8}\nn\\
   &\frac{9 \sqrt{\frac{3}{2}} R_1^4}{8388608\  r^9} Pa_9^{UV}(q_1,R_1,h_1)+\cdots 
   \end{align}
  \noindent where,
   \begin{align}
   &Pa_9^{UV}(q_1,R_1,h_1)=\left(4096 h_1 (3941 + 93312 q1^2 + 
      322560 q_1^4)\right. \nn\\
      &\hspace{.5in}\left.+ 243 (3297681 + 129163840 q_1^2 + 912975360 q_1^4 + 1851260928 q_1^6 + 
  528482304 q_1^8) R_1^6)\right)\nn\\
  \end{align}

 We then have to  analyze if this candidate solution obtained by forward integration has indeed a UV  where the functions are given by 
eq. \eqref{eq:sol-UV} or not. To this end,  we define a mismatch function,
\be\label{eq:mismatch}
m=\sum_i\Big( \log\left(|f_i^{numerical}(r_{match})|\right)- \log\left(|f_i^{expansion}(r)|\right)\Big)^2,
\ee
where $f_i \in \{a,\ b,\ c,\ f\}$, $f_i^{numerical}$ refers to the solution obtained by forward integration and $f_i^{expansion}$ refers to the UV expansion. We then minimize $m$ using \textbf{ NMinimize} and \textbf{ AccuracyGoal}$~=20$ . 
If the minimization procedure yields a small value  ($m\le10^{-4}$) this setup determines the UV parameters $R_1, q_1, h_1$ for which our numerical solution has the required UV behavior. 
 Some sample solutions obtained with this procedure are presented in Figure \ref{fig:sols}. Note that we choose  to normalize the dilaton such that 
\be
(g_s N)^ {3/4} e^{2 \phi_0/ 3} =1 
\ee
where $\phi_0\equiv \phi(r=0)$.

\begin{figure}[p]   
\begin{center}   
	\mbox{   
\subfigure[] {\includegraphics[angle=0,   
width=0.45\textwidth]{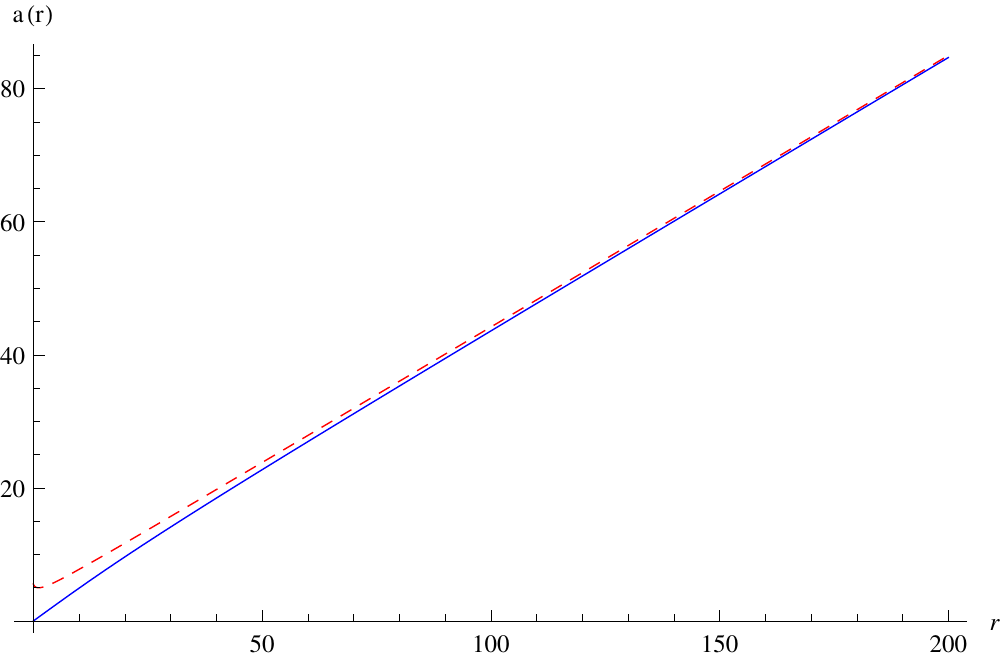} }   
 \subfigure[] {\includegraphics[angle=0,   
width=0.45\textwidth]{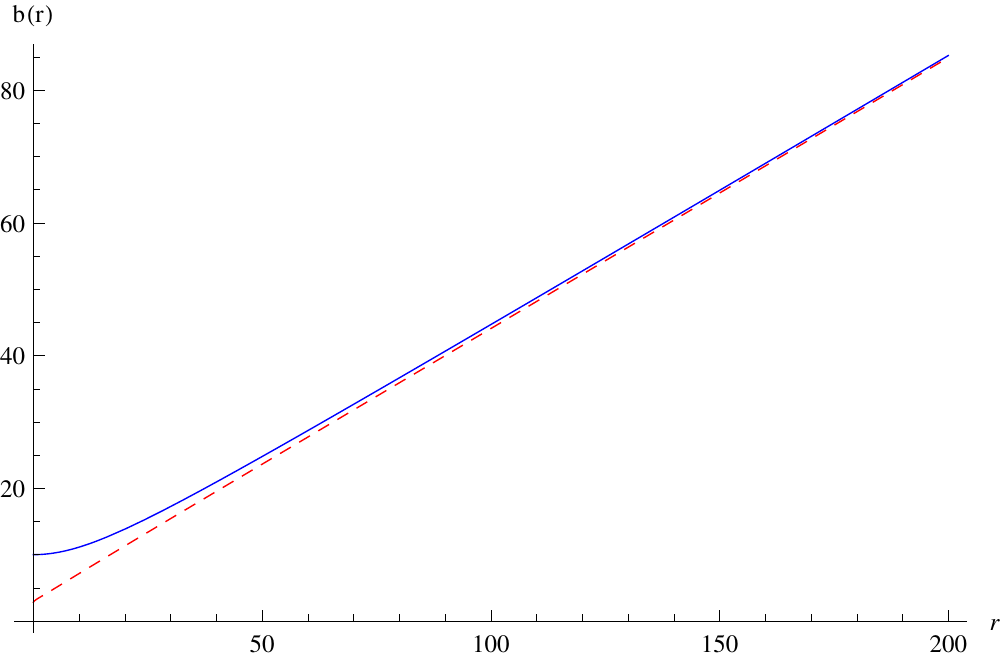} }   
}   
\mbox{   
\subfigure[] {\includegraphics[angle=0,   
width=0.45\textwidth]{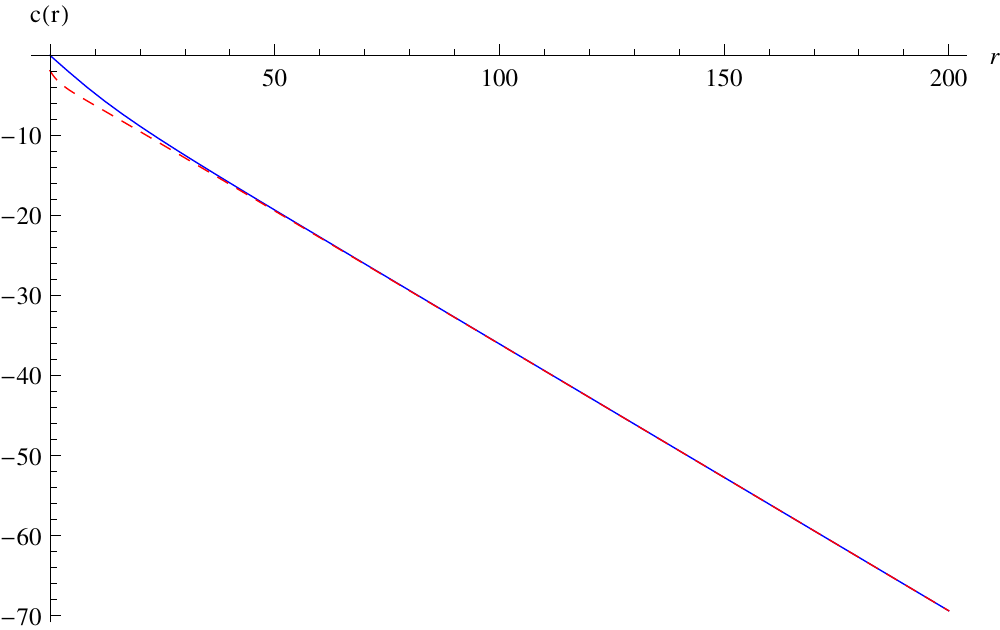} }   
 \subfigure[] {\includegraphics[angle=0,   
width=0.45\textwidth]{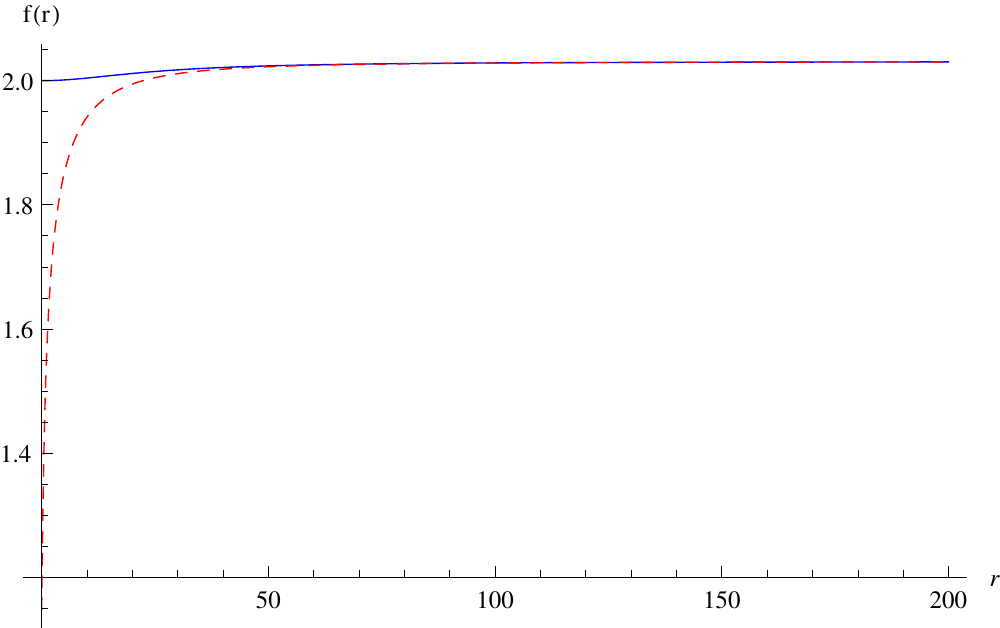} }   
}   
\mbox{   
\subfigure[] {\includegraphics[angle=0,   
width=0.45\textwidth]{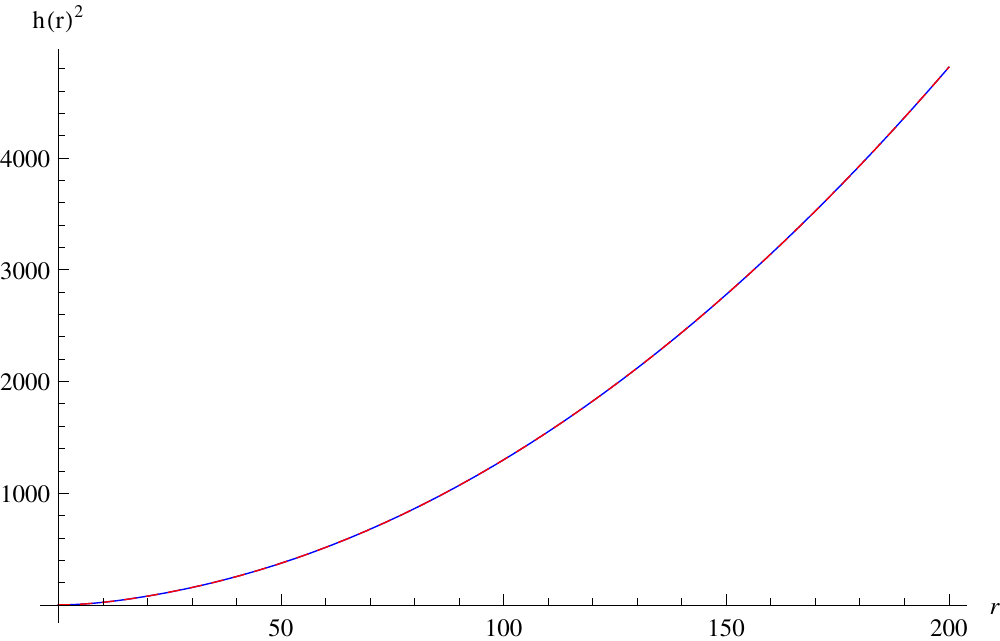} }   
 \subfigure[] {\includegraphics[angle=0,   
width=0.45\textwidth]{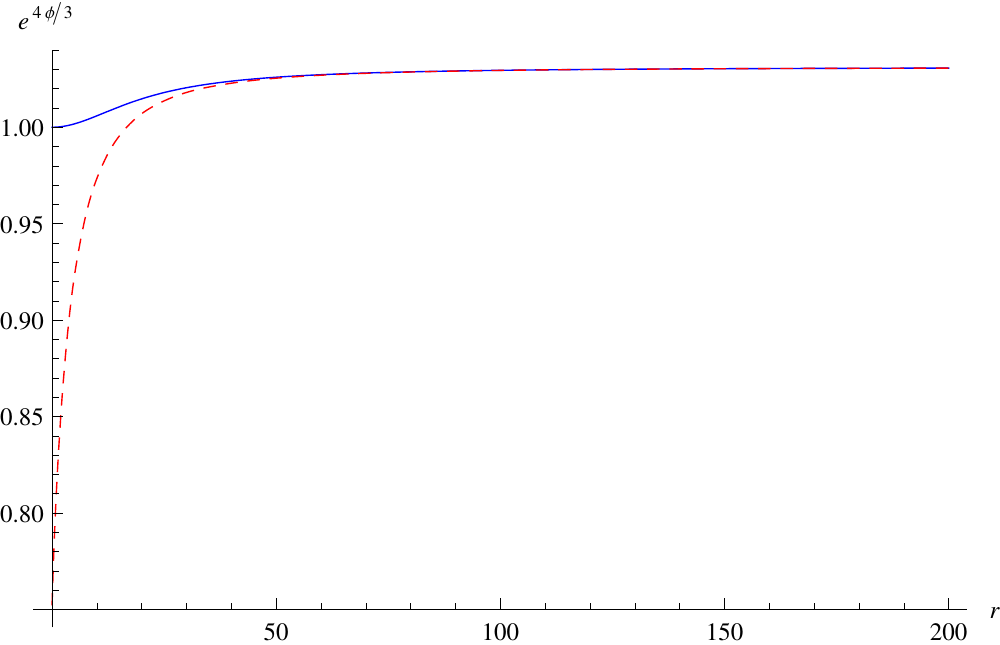} } 
}
\caption{\small  The blue curves are the result of forward integration    
with $R_0=10,\ q_0=1/5 $. After the minimization procedure we obtain the UV parameters    
$q_1 = 1.31946,\  R_1= -2.03087,\  h_1 = -1.9733$ and plot (dashed red lines) the result of   
 integrating back with these parameters to show that it coincides with the forward integration.    
 The small discrepancies in the IR are due to accumulated numerical error. The mismatch function for this solution  is $m < 10^{-4}$. We also plot $h(r)^2$  and $e^{4\phi/3}$ defined in \eqref{metricxxx}}   
\label{fig:sols-app}   
\end{center}   
\end{figure}

\begin{figure}[!ht]   
\begin{center}   
	\mbox{   
\subfigure[] {\includegraphics[angle=0,   
width=0.45\textwidth]{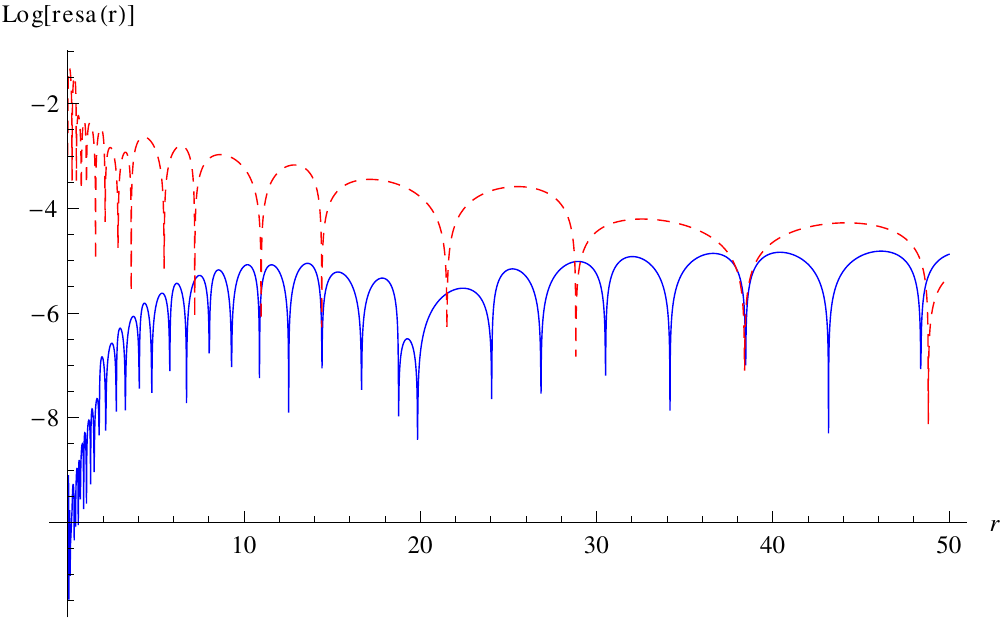} }   
 \subfigure[] {\includegraphics[angle=0,   
width=0.45\textwidth]{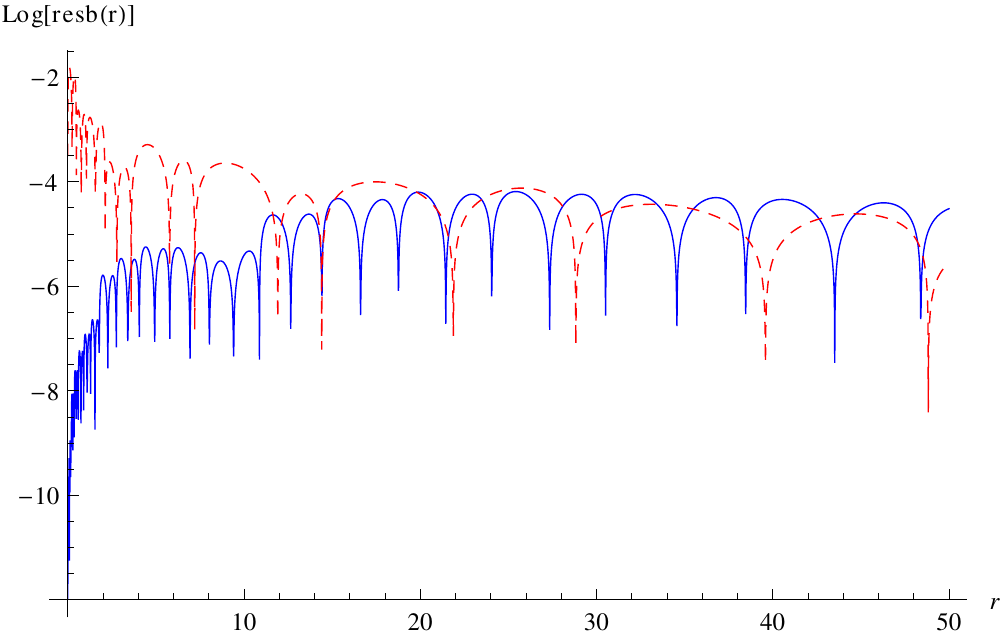} }   
}   
\mbox{   
\subfigure[] {\includegraphics[angle=0,   
width=0.45\textwidth]{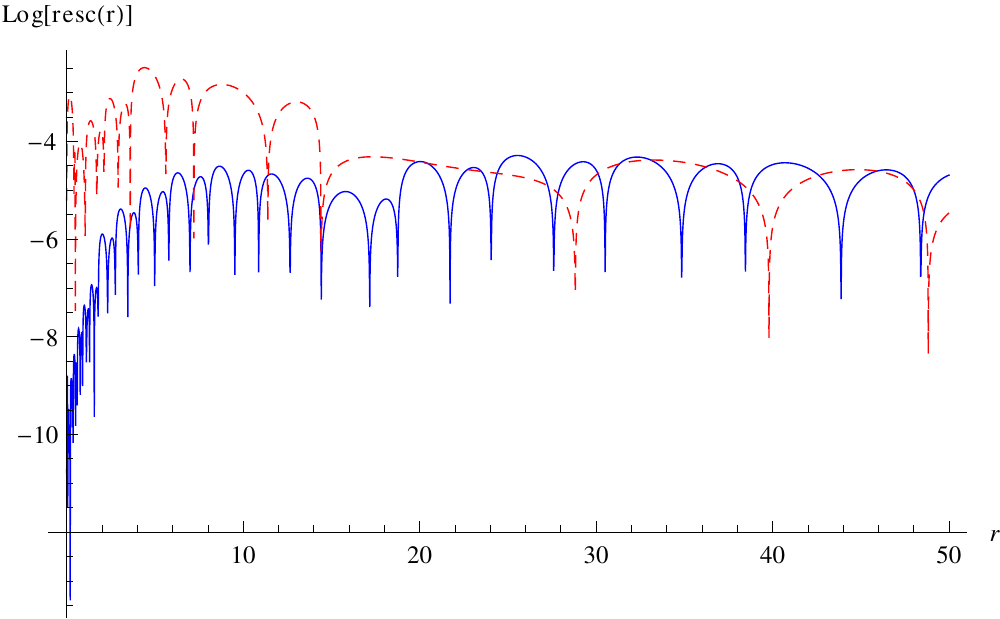} }   
 \subfigure[] {\includegraphics[angle=0,   
width=0.45\textwidth]{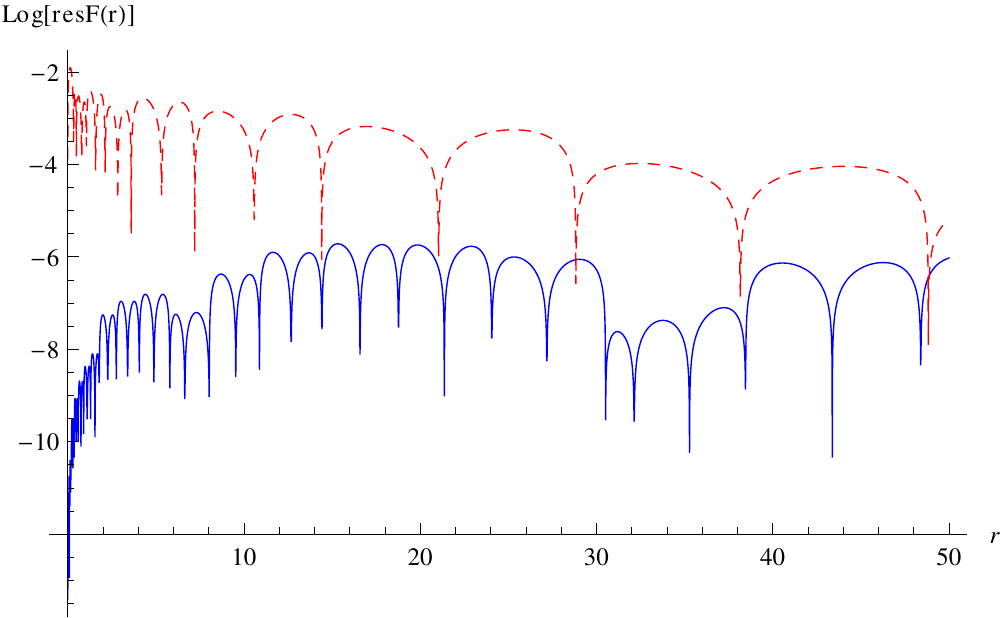} }   
}   
\caption{\small $\log_{10}$ plot of the residuals defined in \eqref{eq:residuals}. The solid blue line is for the  solution obtained by integrating forward (IR to UV) , dashed  line is for the solution obtained by integrating from the UV back to the IR.}\label{fig:res}   
\end{center}   
\end{figure}

A natural question to ask is to what extent  integrating back with the parameters found through the minimization procedure will reproduce the integrated forward solution. Since the IR expansions are of very high order (${\mathcal O}(r^{27})$) while the UV expansions are only of order $\mathcal{O}(1/r^{9})$ we expect that  the UV solution will not be very accurate  in the IR. We present plots comparing the backward and forward integrated solutions in Figure \ref{fig:sols-app}.
In order to verify that the small discrepancies in the IR are due to accumulated numerical error we evaluate the residual. Namely, we define a function  $res_k$ that evaluates the equation of motion for $k(r)$  using the numerical solution. If the solution were exact $res_k$ should be identically zero. Since it is a numerical solution there will always be certain deviation form zero.  
\begin{align}\label{eq:residuals}   
&res_a(r)=\lvert\dot{a}_{num}  +\frac{c_{num}}{2a_{num}} - \frac{a_{num}^5 f_{num}^2}{8 b_{num}^4 c_{num}^3}\lvert, \nn\\   
&res_b(r)=\lvert \dot{b}_{num}  +\frac{c_{num}}{2b_{num}} + \frac{a_{num}^2( a_{num}^2-3c_{num}^2)f_{num}^2}{8b_{num}^3c_{num}^3}\lvert, \nn \\   
&res_c(r)=\lvert \dot{c}_{num}  +1-\frac{c_{num}^2}{2 a_{num}^2}-\frac{c_{num}^2}{2 b_{num}^2}+\frac{3 a_{num}^2   
	f_{num}^2}{8b_{num}^4}\lvert, \nn\\
&res_f(r)=\lvert\dot{f}_{num}  +\frac{a_{num}^4 f_{num}^3}{4 b_{num}^4 c_{num}^3}\lvert.  
\end{align}   
In Figure \ref{fig:res} we see that  the integrated forward solution is  more accurate  for all values of $r$. Also note, (Figure \ref{fig:res} a,  b and d )  that  the integrated back solution fails considerably close to the IR ($resa(r_{IR}) \sim 10^{-2}$) and this explains the differences in figure  \ref{fig:sols-app}.

\begin{thebibliography}{99}   
\bibitem{Maldacena:1997re}    
  J.~M.~Maldacena,   
  Adv.\ Theor.\ Math.\ Phys.\  {\bf 2}, 231 (1998)   
  [hep-th/9711200].   
   
\bibitem{Witten:1998qj}    
  E.~Witten,   
  Adv.\ Theor.\ Math.\ Phys.\  {\bf 2}, 253 (1998)   
  [hep-th/9802150].   
  S.~S.~Gubser, I.~R.~Klebanov and A.~M.~Polyakov,   
  Phys.\ Lett.\ B {\bf 428}, 105 (1998)   
  [hep-th/9802109].   
   
\bibitem{Klebanov:2000hb}    
  I.~R.~Klebanov and M.~J.~Strassler,   
  JHEP {\bf 0008}, 052 (2000)   
  [hep-th/0007191].   
  J.~M.~Maldacena and C.~Nunez,   
  Phys.\ Rev.\ Lett.\  {\bf 86}, 588 (2001)   
  [hep-th/0008001].   
  E.~Witten,   
  Adv.\ Theor.\ Math.\ Phys.\  {\bf 2}, 505 (1998)   
  [hep-th/9803131].   
   
   
\bibitem{Atiyah:2000zz}    
  M.~Atiyah, J.~M.~Maldacena and C.~Vafa,   
  J.\ Math.\ Phys.\  {\bf 42}, 3209 (2001)   
  [hep-th/0011256].   
  B.~S.~Acharya and E.~Witten,   
  hep-th/0109152.   
  M.~Atiyah and E.~Witten,   
  Adv.\ Theor.\ Math.\ Phys.\  {\bf 6}, 1 (2003)   
  [hep-th/0107177].   
   
   
\bibitem{Brandhuber:2001kq}    
  A.~Brandhuber,   
  Nucl.\ Phys.\ B {\bf 629}, 393 (2002)   
  [hep-th/0112113].   
   
\bibitem{Cvetic:2001kp}    
  M.~Cvetic, G.~W.~Gibbons, H.~Lu and C.~N.~Pope,   
  Phys.\ Lett.\ B {\bf 534}, 172 (2002)   
  [hep-th/0112138].   
   
   
   
\bibitem{Brandhuber:2001yi}    
  A.~Brandhuber, J.~Gomis, S.~S.~Gubser and S.~Gukov,   
  Nucl.\ Phys.\ B {\bf 611}, 179 (2001)   
  [hep-th/0106034].   
  M.~Cvetic, G.~W.~Gibbons, H.~Lu and C.~N.~Pope,   
  hep-th/0206154.   
  R.~Hernandez and K.~Sfetsos,
  Phys.\ Lett.\ B {\bf 536}, 294 (2002)
  [hep-th/0202135].
   
\bibitem{Gauntlett:2002sc}    
  J.~P.~Gauntlett, D.~Martelli, S.~Pakis and D.~Waldram,   
  Commun.\ Math.\ Phys.\  {\bf 247}, 421 (2004)   
  [hep-th/0205050].   
  P.~Koerber and D.~Tsimpis,   
  JHEP {\bf 0708}, 082 (2007)   
  [arXiv:0706.1244 [hep-th]].   

\bibitem{Grana:2004bg}    
  M.~Grana, R.~Minasian, M.~Petrini and A.~Tomasiello,   
  JHEP {\bf 0408}, 046 (2004)   
  [hep-th/0406137].   
  L.~Martucci and P.~Smyth,   
  JHEP {\bf 0511}, 048 (2005)   
  [hep-th/0507099].   
   
   
   
\bibitem{Maldacena:2009mw}    
  J.~Maldacena and D.~Martelli,   
  JHEP {\bf 1001}, 104 (2010)   
  [arXiv:0906.0591 [hep-th]].   
  J.~Gaillard, D.~Martelli, C.~Nunez and I.~Papadimitriou,   
  Nucl.\ Phys.\ B {\bf 843}, 1 (2011)   
  [arXiv:1004.4638 [hep-th]].   
  E.~Caceres, C.~Nunez and L.~A.~Pando-Zayas,
  JHEP {\bf 1103}, 054 (2011)
  [arXiv:1101.4123 [hep-th]].
  E.~Caceres and S.~Young,
  Phys.\ Rev.\ D {\bf 87}, no. 4, 046006 (2013)
  [arXiv:1205.2397 [hep-th]].
  S.~Bennett, E.~Caceres, C.~Nunez, D.~Schofield and S.~Young,
  JHEP {\bf 1205}, 031 (2012)
  [arXiv:1111.1727 [hep-th]].
  D.~Elander, J.~Gaillard, C.~Nunez and M.~Piai,   
  JHEP {\bf 1107}, 056 (2011)   
  [arXiv:1104.3963 [hep-th]].   
  E.~Conde, J.~Gaillard, C.~Nunez, M.~Piai and A.~V.~Ramallo,   
  JHEP {\bf 1202}, 145 (2012)   
  [arXiv:1112.3350 [hep-th]].   
   
\bibitem{de la Ossa:1992vc}    
  X.~C. de~la~Ossa and F.~Quevedo,   
  Nucl.\ Phys.\ B {\bf 403}, 377 (1993)   
  [hep-th/9210021].   
  E.~Alvarez, L.~Alvarez-Gaume, J.~L.~F.~Barbon and Y.~Lozano,   
  Nucl.\ Phys.\ B {\bf 415}, 71 (1994)   
  [hep-th/9309039].   
  A.~Giveon and M.~Rocek,   
  Nucl.\ Phys.\ B {\bf 421}, 173 (1994)   
  [hep-th/9308154].   
  Y.~Lozano,   
  Phys.\ Lett.\ B {\bf 355}, 165 (1995)   
  [hep-th/9503045].   
  K.~Sfetsos,
  Phys.\ Rev.\ D {\bf 50}, 2784 (1994)
  [hep-th/9402031].
	
\bibitem{Sfetsos:2010uq}    
  K.~Sfetsos and D.~C.~Thompson,   
  Nucl.\ Phys.\ B {\bf 846}, 21 (2011)   
  [arXiv:1012.1320 [hep-th]].   
  Y.~Lozano, E.~.O Colgain, K.~Sfetsos and D.~C.~Thompson,   
  JHEP {\bf 1106}, 106 (2011)   
  [arXiv:1104.5196 [hep-th]].   
   
\bibitem{Itsios:2013wd}    
  G.~Itsios, C.~Nunez, K.~Sfetsos and D.~C.~Thompson,   
  Nucl.\ Phys.\ B {\bf 873}, 1 (2013)   
  [arXiv:1301.6755 [hep-th]].   
   
\bibitem{Lozano:2012au}    
  Y.~Lozano, E.~OColgain, D.~Rodriguez-Gomez and K.~Sfetsos,   
  Phys.\ Rev.\ Lett.\  {\bf 110}, 231601 (2013)   
  [arXiv:1212.1043 [hep-th]].   
  G.~Itsios, C.~Nunez, K.~Sfetsos and D.~C.~Thompson,   
  Phys.\ Lett.\ B {\bf 721}, 342 (2013)   
  [arXiv:1212.4840].   
  N.~T.~Macpherson,   
  arXiv:1310.1609 [hep-th].   
  E.~Gevorgyan and G.~Sarkissian,   
  arXiv:1310.1264 [hep-th].   
  Y.~Lozano, E.~OColgain and D.~Rodriguez-Gomez,
  arXiv:1311.4842 [hep-th].
  S.~Zacarías,
  arXiv:1401.7618 [hep-th].

\bibitem{Sfetsos:2013wia} 
  K.~Sfetsos,
  arXiv:1312.4560 [hep-th].

   
   
\bibitem{Kaste:2002xs}   
  P.~Kaste, R.~Minasian, M.~Petrini and A.~Tomasiello,   
  JHEP {\bf 0209} (2002) 033   
  [hep-th/0206213].   
   
\bibitem{Gaillard:2009kz}   
  J.~Gaillard and J.~Schmude,   
  JHEP {\bf 1002} (2010) 032   
  [arXiv:0908.0305 [hep-th]].   
   
   
   
\bibitem{Edelstein:2001pu}   
  J.~D.~Edelstein and C.~Nunez,   
  JHEP {\bf 0104}, 028 (2001)   
  [hep-th/0103167].   


\bibitem{Andriot:2008va}
  D.~Andriot,
  JHEP {\bf 0808} (2008) 096
  [arXiv:0804.1769 [hep-th]].

\bibitem{andriotxx}
   R.~Minasian, M.~Petrini and A.~Zaffaroni,
Geometry,''
   JHEP {\bf 0612}, 055 (2006)
   [hep-th/0606257].
   A.~Butti, D.~Forcella, L.~Martucci, R.~Minasian, M.~Petrini and 
A.~Zaffaroni,
theories,''
   JHEP {\bf 0807}, 053 (2008)
   [arXiv:0712.1215 [hep-th]].
   J.~McOrist, D.~R.~Morrison and S.~Sethi,
   Adv.\ Theor.\ Math.\ Phys.\  {\bf 14} (2010)
   [arXiv:1004.5447 [hep-th]].

\bibitem{Gaillard:2013vsa}
  J.~Gaillard, N.~T.~Macpherson, C.~Nunez and D.~C.~Thompson,
  arXiv:1312.4945 [hep-th].
	
\bibitem{Andriot:2010sya}
   D.~Andriot,
   ``String theory flux vacua on twisted tori and Generalized Complex 
Geometry,''. PhD. thesis.


\bibitem{Petrini:2009ur} 
  M.~Petrini and A.~Zaffaroni,
  JHEP {\bf 0909}, 107 (2009)
  [arXiv:0904.4915 [hep-th]].

	
\bibitem{Barranco:2013fza}
  A.~Barranco, J.~Gaillard, N.~T.~Macpherson, C.~Nunez and D.~C.~Thompson,
  JHEP {\bf 1308} (2013) 018
  [arXiv:1305.7229 [hep-th]].

\bibitem{Itzhaki:1998dd}    
  N.~Itzhaki, J.~M.~Maldacena, J.~Sonnenschein and S.~Yankielowicz,   
  Phys.\ Rev.\ D {\bf 58}, 046004 (1998)   
  [hep-th/9802042].   

   
   
\bibitem{Nunez:2009da}   
  C.~Nunez, M.~Piai and A.~Rago,   
  Phys.\ Rev.\ D {\bf 81}, 086001 (2010)   
  [arXiv:0909.0748 [hep-th]].   
\bibitem{Acharya:2001dz}   
  B.~S.~Acharya and C.~Vafa,   
  hep-th/0103011.   
   
   
\bibitem{Gursoy:2003hf}    
  U.~Gursoy, S.~A.~Hartnoll and R.~Portugues,   
  Phys.\ Rev.\ D {\bf 69}, 086003 (2004)   
  [hep-th/0311088].   
   
   
   
   
   
   
\bibitem{Girardello:1998pd}    
  L.~Girardello, M.~Petrini, M.~Porrati and A.~Zaffaroni,   
  JHEP {\bf 9812}, 022 (1998)   
  [hep-th/9810126].   
   
\bibitem{Ryu:2006bv}    
  S.~Ryu and T.~Takayanagi,   
  Phys.\ Rev.\ Lett.\  {\bf 96}, 181602 (2006)   
  [hep-th/0603001].   
   
\bibitem{Klebanov:2007ws}    
  I.~R.~Klebanov, D.~Kutasov and A.~Murugan,   
  Nucl.\ Phys.\ B {\bf 796}, 274 (2008)   
  [arXiv:0709.2140 [hep-th]].   
   
   
   
   
\bibitem{Kol:2014nqa} 
  U.~Kol, C.~Nunez, D.~Schofield, J.~Sonnenschein and M.~Warschawski,
  JHEP {\bf 1406}, 005 (2014)
  [arXiv:1403.2721 [hep-th]].
   
\bibitem{Barbon:2008ut}    
  J.~L.~F.~Barbon and C.~A.~Fuertes,   
  JHEP {\bf 0804}, 096 (2008)   
  [arXiv:0803.1928 [hep-th]].   
   
\bibitem{Marolf:2000cb}   
  D.~Marolf,   
  hep-th/0006117.   
   
   
\bibitem{Benini:2007gx}    
  F.~Benini, F.~Canoura, S.~Cremonesi, C.~Nunez and A.~V.~Ramallo,   
  JHEP {\bf 0709}, 109 (2007)   
  [arXiv:0706.1238 [hep-th]].   
  F.~Benini, F.~Canoura, S.~Cremonesi, C.~Nunez and A.~V.~Ramallo,
  JHEP {\bf 0702}, 090 (2007)
  [hep-th/0612118].
   
   
\bibitem{nuevos}
Yolanda Lozano, Niall Macpherson and Eoin O'Colgain. To appear.
Niall Macpherson, Carlos Nunez, Leo Pando-Zayas, Vincent Rodgers, 
Catherine Withing. To appear.
  
  
\end{thebibliography}
\end{document}